\documentclass[twocolumn,secnumarabic,amssymb, nobibnotes, aps, prd,superscriptaddress,showkeys]{revtex4-2}
\usepackage{graphicx}
\usepackage{multirow}
\usepackage{amsmath,amssymb,amsfonts}
\usepackage{mathrsfs}
\usepackage{xcolor}
\usepackage{textcomp}
\usepackage{algorithm}
\usepackage{algorithmicx}
\usepackage{algpseudocode}
\usepackage{listings}
\usepackage{caption}
\usepackage{float}
\usepackage{subcaption}
\usepackage{pifont}
\usepackage{tabularx}
\usepackage{hyperref}

\definecolor{eye-caring}{RGB}{220,250,220}

\newcommand{\etal}{\emph{et al.} \,}

\begin{document}

\title{Learning the expressibility of quantum circuit ansatz using transformer}

\author{Fei Zhang}
\email{zhangfei@htu.edu.cn}
\affiliation{College of Computer and Information Engineering, Henan Normal University, Xinxiang, China}
\affiliation{Key Laboratory of Artificial Intelligence and Personalized Learning in Education of Henan Province}
\author{Jie Li}
\affiliation{College of Computer and Information Engineering, Henan Normal University, Xinxiang, China}
\author{Zhimin He}
\affiliation{School of Electronic and Information Engineering, Foshan University, Foshan 528000, China}
\author{Haozhen Situ}
\email{situhaozhen@gmail.com}
\affiliation{College of Mathematics and Informatics, South China Agricultural University, Guangzhou 510642, China}

\begin{abstract}
	
With the exponentially faster computation for certain problems, quantum computing has garnered significant attention in recent years. Variational quantum algorithms are crucial methods to implement quantum computing, and an appropriate task-specific quantum circuit ansatz can effectively enhance the quantum advantage of VQAs. However, the vast search space makes it challenging to find the optimal task-specific ansatz. Expressibility, quantifying the diversity of quantum circuit ansatz states to explore the Hilbert space effectively, can be used to evaluate whether one ansatz is superior to another. In this work, we propose using a transformer model to predict the expressibility of quantum circuit ansatze. We construct a dataset containing random PQCs generated by the gatewise pipeline, with varying numbers of qubits and gates. The expressibility of the circuits is calculated using three measures: KL divergence, relative KL divergence, and maximum mean discrepancy. A transformer model is trained on the dataset to capture the intricate relationships between circuit characteristics and expressibility. Four evaluation metrics are employed to assess the performance of the transformer. Numerical results demonstrate that the trained model achieves high performance and robustness across various expressibility measures. This research can enhance the understanding of the expressibility of quantum circuit ansatze and advance quantum architecture search algorithms.

\end{abstract}


\maketitle

\section{Introduction}\label{sec1}

The principles of quantum superposition and interference enable quantum computing to be widely applied in areas where traditional computing is ineffective, such as quantum simulation \cite{endo2020variational}. For instance, in the field of biochemical pharmaceuticals,  the introduction of quantum technology can accelerate the development of new drugs and facilitate personalized medicine \cite{QuBiotechnology2022}. Variational quantum algorithm (VQA) is a popular method for exploring  quantum advantage in the noisy intermediate-scale quantum (NISQ) era \cite{cerezo2021variational} .  
VQA achieves the optimal value of the objective function for a given task by iteratively optimizing the parameters of quantum gates in the quantum circuit ansatz, commonly referred to as parameterized quantum circuits (PQCs) in the literature \cite{situ2020quantum,pan2023deep,ding2023active,shi2023parameterized,ni2024multilevel}. 
Selecting an appropriate ansatz for a given task is crucial for simplifying the optimization process and achieving optimal values. Quantum architecture search (QAS) aims to automatically identify the optimal  circuit structure for a given task \cite{martyniuk2024quantum}. However, the vast search space and the time-consuming nature of training circuits to evaluate their actual performance present significant  challenges. Recent research has focused on evaluating circuit structures without optimizing gate parameters \cite{anagolum2024elivagar,zhimin2024aaai,situ2024distributed}, with expressibility serving as a useful proxy for assessing PQC performance.

PQCs with stronger diversity in generated quantum states imply a greater ability to explore the Hilbert space. Sim \etal proposed using the Kullback-Leibler (KL) divergence between the fidelity distribution of states sampled from a PQC with random parameters and the fidelity distribution of Haar random states to evaluate the expressibility of a PQC \cite{HybridExpressibility2019}. Rasmussen \etal introduced relative expressibility, a normalized measure of a circuit's expressibility compared to an idle circuit \cite{KL_R2020}. This is calculated by taking the negative logarithm of the expressibility ratio to ensure a positive value. To accommodate noisy environments, Ding \etal proposed an expressibility measure based on the maximum mean discrepancy (MMD) between the output distribution of a PQC and the uniform distribution \cite{ExpMMD2022}. Calculating the expressibility of a PQC involves obtaining multiple quantum states produced by the circuit. For example, Sim \etal sampled 5000 quantum state fidelities from a single PQC to construct a histogram for calculating KL divergence  \cite{HybridExpressibility2019}. This process becomes highly time-consuming when applied to a large number of PQCs. Therefore, employing deep learning techniques to estimate the expressibility of PQCs is crucial. Such estimation methods can greatly benefit the study of PQCs, including QAS.

\begin{figure*}[!htbp]
	\centering
	\includegraphics[width=0.8\linewidth]{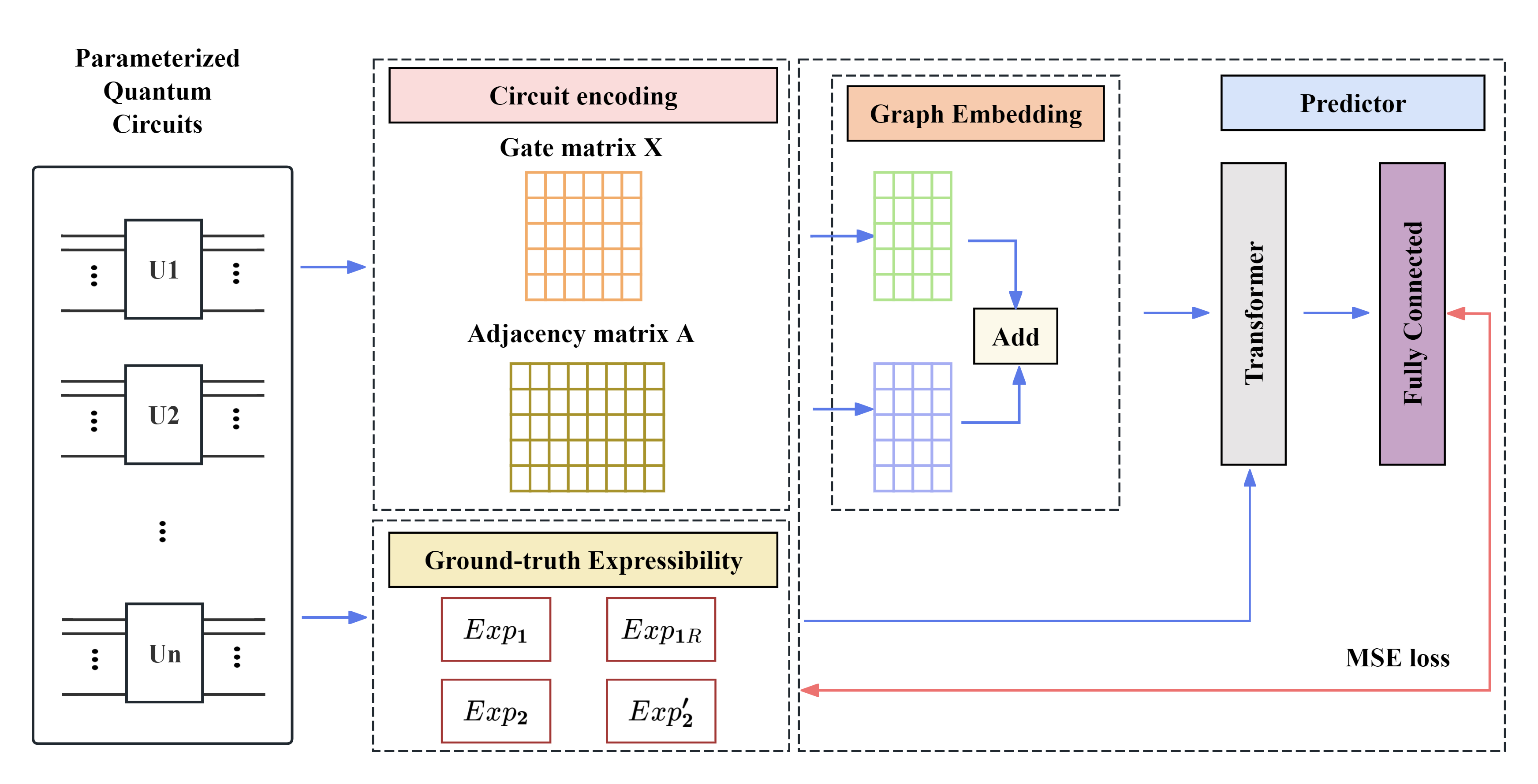} 
	\caption{Framework of the proposed expressibility estimation method. The method comprises three phases: (1) graph encoding of the PQC, (2) ground-truth expressibility calculation, and (3) transformer model training.}
	\label{fig:framework} 
\end{figure*}

In this work, we present an approach that converts PQCs into graphs and estimates their expressibility using a transformer model. Figure \ref{fig:framework} illustrates our expressibility estimation framework. Initially, we generate random circuits with varying numbers of qubits and gates using the gatewise pipeline \cite{zhangHsieh2021}.  These circuits are then converted into graphs, where each quantum gate is represented as a node, and directed edges indicate the precedence relationships between gates. We construct a gate matrix by extracting features for each node based on the type of quantum gate and the qubit(s) it acts on, and we use an adjacency matrix to represent the relationships between nodes. For these circuits, expressibility is calculated using three measures: KL divergence \cite{HybridExpressibility2019}, relative KL divergence \cite{KL_R2020}, and MMD \cite{ExpMMD2022}. To further investigate the effect of noise, depolarizing and bit-flip noise are introduced to these circuits, and expressibility is then calculated solely using MMD. We then employ a transformer model to explore the relationship between the stuctures of PQCs and their expressibility. The model is trained using the node features and the adjacency matrix to minimize the difference between the predicted and the actual expressibility. Numerical results demonstrate that our method performs well across four evaluation metrics: root mean square error (RMSE), $R^2$, Spearman correlation coefficient, and Kendall correlation coefficient, both in the presence and absence of noise.

The contributions of this work are summarized as follows:

\begin{enumerate}
	\item[(1)] Random PQCs with varying numbers of qubits and gates are constructed. For each qubit count ranging from 4 to 6, 10000 random circuit samples are generated. For each circuit sample, we calculate the expressibility using KL divergence, relative KL divergence, and MMD, resulting in three labels for expressibility prediction. Additionally, MMD is used to calculate the expressibility of each circuit under the influence of noise, providing an extra label.  
	\item[(2)] The transformer model is trained on the complete set of circuit samples to predict the expressibility of PQCs. The numerical results demonstrate that the RMSEs between the predicted and actual expressibility values are very small. Additionally, the Spearman and Kendall correlation coefficients demonstrate a strong correlation between the predicted and actual expressibility. These observations suggest that the transformer model is capable of providing highly reiable expressibility estimations. 
        \item[(3)] We open-source the dataset and the transformer model to support research of PQCs and QAS. The database created in this paper, as well as the source code of the proposed method, can be found on GitHub at \href{https://github.com/FeiZhang-Y/Quantum_architecture_search/}{https://github.com/FeiZhang-Y/Quantum\_architecture\_search/}.

\end{enumerate}

The remainder of the paper is organized as follows: Section \ref{RelatedWork} offers a brief overview of expressibility calculation, quantum circuit encoding, and the transformer model. Section \ref{ProMethodData} details the construction of the dataset for circuit expressibility estimation and explores the relationship between circuit expressibility and circuit properties. Section \ref{ProMethod} describes the construction of the transformer model used in this work. Section \ref{Experiment} outlines the numerical experiment setup and presents the results. Finally, Section \ref{Conclusion} summarizes the key findings and conclusions of the study.

\section{Related Work}\label{RelatedWork}

\subsection{Expressibility measures}

The expressibility based on the KL divergence 
\begin{equation}
	\begin{split}
		Exp_1  &=  D_{KL} (\hat{P}_{PQC}(F;\theta) || P_{Haar}(F))  \\		
		&= \sum_{F}  \hat{P}_{PQC}(F;\theta) ln \frac{\hat{P}_{PQC}(F;\theta)}{P_{Haar}(F)} 
	\end{split}
\end{equation}
has been introduced as a measure for quantifying the disparity between the distribution of state fidelities generated by a PQC and that generated by Haar random states \cite{HybridExpressibility2019}. A smaller $Exp_1$ value signifies that a PQC can generate more diverse states in Hilbert space, indicating better expressibility. Suppose a state generated by a PQC is denoted as $U(\theta)|0 \rangle$, with the parameter $\theta$ randomly drawn from a uniform distribution over  $[0,2\pi]$. The fidelity $F$ between two generated states $|\varphi_1\rangle$ and $|\varphi_2\rangle$ is $| \langle \varphi_1|\varphi_2 \rangle|^2$. $\hat{P}_{PQC}(F;\theta)$ denotes the distribution of state fidelities generated by a PQC. Meanwhile, $P_{Haar}(F)$ represents the distribution of state fidelities produced by $N$-qubit Haar random states, characterized by an analytical form of $(2^N-1)(1-F)^{2^N-2}$.  

The calculation of $Exp_1$ depends on the number of histogram bins ($n_{\text{bins}}$) used to partition the interval $[0,1]$. Rasmussen \etal introduced the relative expressibility 
\begin{equation}
Exp_{1R} = -\ln \frac{Exp_1}{Exp_1(\text{Idle})}
\end{equation}
to standardize $Exp_1$ against the idle circuit utilizing the Identity gate \cite{KL_R2020}. $Exp_1(\text{Idle})$ is given by $(2^N-1) \ln n_{\text{bins}}$. To enhance the distinction between the most expressive circuits ($Exp_1 \approx 0$) and make the result positive, a negative logarithm function is applied. Therefore, a higher $Exp_{1R}$ value signifies greater circuit expressibility. 

However, both $Exp_1$ and $Exp_{1R} $ are designed for PQCs that generate pure states. In order to investigate the impact of noise on PQCs, Ding \etal \cite{ExpMMD2022} proposed an expressibility measure
\begin{equation}
\begin{aligned}
& Exp_2 = 1-MMD(F_U, F_{\text{uniform}})  \\
& \approx 1 \!- \! \frac{1}{M^2} \left | \sum_{i=1,j=1}^{M} \!\!\! k(X_i,X_j) \! + \! k(Y_i,Y_j) \!-\! 2k(X_i,Y_j) \right|,
\end{aligned}
\end{equation}
which utilizes the maximum mean discrepancy (MMD) to quantify the disparity between the output distribution of a PQC ($F_U$) and the uniform distribution ($F_{\text{uniform}}$) in the simplex 
\begin{align}
\Delta^{2^N} = \{ (p_0, p_1, \ldots, p_{2^N-2} | \sum_i p_i < 1, p_i \geq  0).    
\end{align}
A larger value of $Exp_2$ reflects better expressibility of the PQC, with $Exp_2=1$ if and only if $F_U=F_{\text{uniform}}$.
Here, $X_i$ and $Y_i$ ($i =1, 2, \cdots, M$) represent samples drawn from the distributions $F_U$ and $F_{\text{uniform}}$, respectively. A Gaussian kernel function $k(x,y) = e^{-\frac{||x-y||^2}{4 \sigma}}$ is utilized to map samples $x$ and $y$ into a higher-dimensional Hilbert space and subsequently calculate the distance between them.  The hyper-parameter $\sigma$  fixed at 0.01 to ensure consistency with the original method. 

In this section, three expressibility measures for PQCs are reviewed.  Their distinctions are summarized in Table \ref{Tab:Exp}. $Exp_1$ and $Exp_{1R}$  assess the expressibility of a PQC by assuming a pure output state with an initial state of $|0\rangle^{\otimes N}$. In contrast, $Exp_2$ can be applied to both noiseless and noisy circuits, with the initial state being $|+ \rangle^{\otimes N}$.

\begin{table}[htp]
	\caption{Distinctions between three expressibility measures}
	\begin{center}
		\resizebox{\linewidth}{!}{
		\begin{tabular}{|l|c|c|c|}
			\hline
			 & $Exp_1$ & $Exp_{1R}$  & $Exp_2$  \\ \hline
			Initial state & $|0 \rangle^{\otimes N}$ & $|0 \rangle^{\otimes N}$ & $|+ \rangle^{\otimes N}$ \\ \hline
			Applicable for mixed states & \ding{56} & \ding{56} & \ding{52}  \\ \hline
			Higher-dimensional mapping & \ding{56} & \ding{56} & \ding{52}  \\ \hline			
			Distance metric & KL & KL & MMD  \\ \hline
			Better expressibility & $\downarrow$ & $\uparrow$ & $\uparrow$ \\ \hline
		\end{tabular}}
	\end{center}
	\label{Tab:Exp}
\end{table}

\subsection{Quantum Circuit Encoding} \label{Encoding}

Encoding quantum circuit is an important preprocessing step for the estimation of circuit expressibility using  transformers. Zhang \etal proposed a one-to-one mapping method to convert the circuit into an image, where the image's height corresponds to the number of qubits, and its width corresponds to the number of layers, with gate types represented as pixels \cite{zhangHsieh2021}. However, this image representation struggles with cases where where the two-qubit gates act on non-adjacent qubits. Mao \etal treated the layers of circuits as temporal sequences and used long short-term memory (LSTM) networks to capture dependencies between gates, but this encoding scheme is sensitive to the position of the qubits \cite{Mao2023zkt}. Altares \etal employed binary encoding for the gate types and rotation angles of parameterized gates, representing circuits as binary strings \cite{altares2021automatic}. Nonetheless, this method cannot handle arbitrary rotation angles. He \etal  encoded circuits as directed acyclic graphs (DAGs) to preserve the topology information of the circuits \cite{GSQAS2023}. DAGs can effectively represent the global structure of circuits and the relationships between gates. Therefore, this work adopts graph encoding to represent the circuits.

\subsection{Transformer model}

The transformer model has achieved noticeable results in natural language processing, computer vision, recommendation systems, multimodal learning, and other fields due to its excellent global information capture capabilities. By introducing the self-attention module and calculating the similarity between the query and key vectors, the transformer can assign higher weights to value vectors when the query and key vectors are highly correlated. This allows the transformer to capture long-distance dependencies between input sequences. Additionally, the multi-head attention mechanism enables the extraction of diverse features, thereby enhancing the representation ability of the transformer \cite{introductiontransformers2023}. Due to the excellent performance of the transformer model in processing sequential data, it has been increasingly applied in the field of quantum computing. For example, transformer is employed to generate ``realistic-looking'' quantum circuits \cite{apak2024ketgpt}, represent the probability distributions of quantum states \cite{Zhang2022TransformerQS}, and estimate circuit reliability through graph-based approaches \cite{QuEst2022}. Additionally, a quantum transformer is designed to address unsupervised visual clustering tasks \cite{nguyen2024qclusformer}. These studies highlight the versatility and potential of transformer in advancing quantum computing.

\section{Circuit Expressibility Dataset} \label{ProMethodData}

Training the transformer model to accurately estimate PQC expressibility requires a dataset encompassing diverse PQCs, each exhibiting various characteristics such as different numbers of qubits, gates, and expressibilities. In this section, we describe the construction of our dataset. Our dataset comprises 30000 PQCs along with their respective $Exp_1$, $Exp_{1R}$, $Exp_{2}$ and $Exp'_{2}$. Here, $Exp_{2}$ and $Exp'_{2}$ represent the MMD-based expressibility of noiseless and noisy circuits, respectively. $Exp_1$ and $Exp_{1R}$ are calculated only for noiseless circuits.

\subsection{Quantum Circuit Generation}

We adopt the gatewise pipeline \cite{zhangHsieh2021} to generate random circuits using parameterized single-qubit gate U3 and two-qubit gate CZ. Specifically, a probability vector from a Gaussian distribution $\mathcal N(0,1.35)$ is generated to select the gate types, and another vector following a  normal distribution $\mathcal N(0,1)$ is generated to determine the gate position. The first layer of the circuit assigns U3 gates to each qubit, because $CZ |00 \rangle  = |00 \rangle$. A CZ gate can only be placed on adjacent qubits, and the qubit connectivity follows a ring topology.

To better analyze the relationship between the characteristics of quantum circuits and their expressibilities, as well as to effectively evaluate the generalization ability of the learning model, it is crucial for the generated circuits to exhibit significant diversity. This diversity is essential to capture the potential complexity and nonlinearity in the relationship between circuit characteristics and expressibility. 

We generate random PQCs with 4, 5 and 6 qubits. To enhance the diversity of PQCs, 20 different gate counts are selected for each qubit count. For a given number of qubits and gates, 500 circuits are randomly generated. Consequently, a total of 10000 circuits are generated for each qubit count. The dataset contain 30000 circuits in total. 

Detailed information regarding the dataset is presented in Table \ref{Tab:data}. The first column shows the adopted gate counts. The second column lists the selected gate counts for each qubit count, which increase with the qubit count. The third column provides the range of depths for the generated circuits for each qubit count. The fourth column indicates the range of the number of U3 gates in the generated circuits for each qubit count.

\begin{table}[htp]
	\caption{Information of the generated dataset}
	\begin{center}
		\begin{tabular}{|c|c|c|c|}
			\hline
			$\#$qubit & $\#$gate & depth & $\#$U3 gate \\ \hline
			4 & 10 $\sim$ 29 & 4 $\sim$ 22 & 4 $\sim$ 20 \\ \hline
			5 & 15 $\sim$ 34 & 5 $\sim$ 22 & 5 $\sim$ 24  \\ \hline
			6 & 20 $\sim$ 39 & 6 $\sim$ 23 & 6 $\sim$ 28 \\ \hline
		\end{tabular}
	\end{center}
	\label{Tab:data}
\end{table}

The expressibility measures $Exp_1$, $Exp_{1R}$ and $Exp_{2}$ are evaluated for each PQC in the dataset, and these measures are treated as target values during the training process. To examine the influence of noise on the expressibility of PQCs, we transform each PQC in the dataset to a noisy version. Specifically, we introduce depolarizing noise after each gate application. The noise strength is set to 0.001 for the U3 gate and 0.01 for the CZ gate. Additionally, bit-flip noise with a strength of 0.01 is added before the measurement to account for readout noise. The expressibility measure $Exp_2$ is then evaluated for each noisy PQC, with the results denoted as $Exp'_2$ to differentiate from the noiseless case.

\subsection{Graph Encoding of Quantum Circuit}\label{GraphEncoding}

Directed acyclic graphs (DAGs) can preserve the topology information of PQCs \cite{GSQAS2023}. Therefore, in this work, DAGs are utilized to encode the circuits. In this representation, the gates are denoted as nodes within the graph.  A directed edge from node $a$ to node $b$ indicates that the qubit under influence by gate $a$ is subsequently influenced by gate $b$. Two special nodes, named ``Start'' and ``End'',  are introduced to represent the input and output of the circuit, respectively. Edges are created from the Start node to the nodes corresponding to the first gates on each qubit, and from the nodes corresponding to the last gates on each qubit to the End node. 

A gate matrix is constructed to represent the feature vectors of the nodes, which are based on the type of quantum gate and the position of target qubit(s).  Meanwhile, an adjacency matrix is constructed to describe the relationships between the nodes.

\subsection{Dataset Properties}

\begin{figure*}[!htbp]
	\centering
 	\subfloat{\includegraphics[width=0.35 \linewidth]{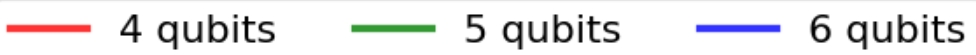}} 
        \vspace{3mm}
  \setcounter{subfigure}{0}
	\subfloat{\includegraphics[width=0.3\linewidth]{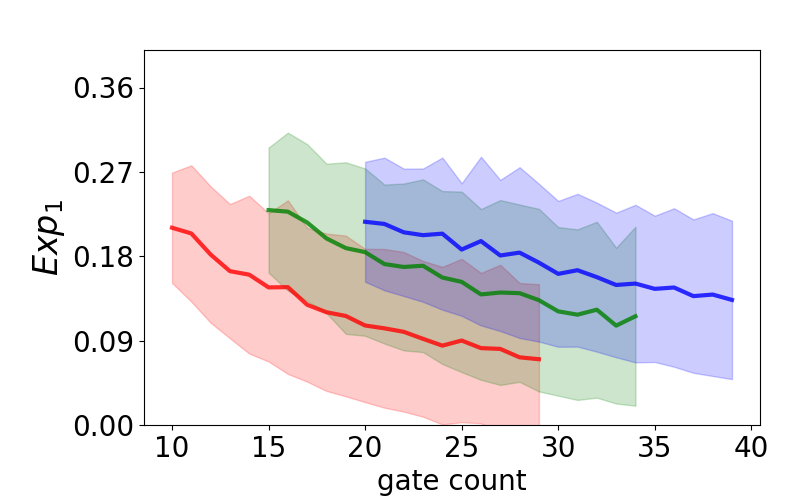}  	
		\label{fig:Exp1_gateNo} 	}
	\subfloat{\includegraphics[width=0.3\linewidth]{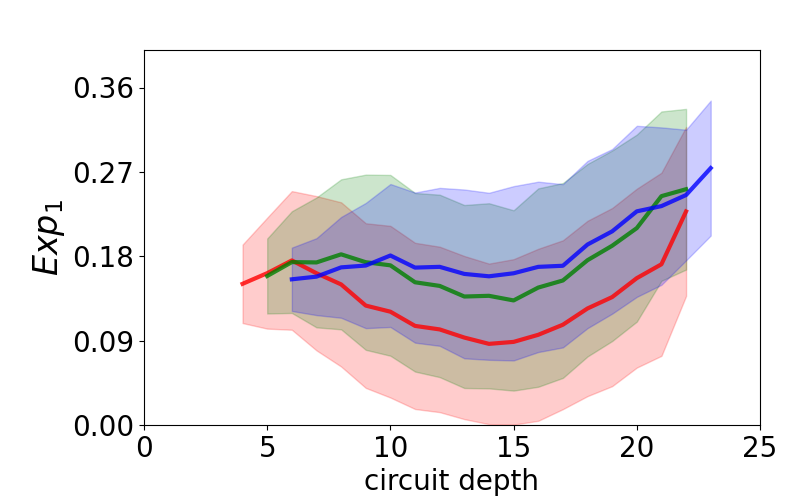} 	\label{fig:Exp1_depth}	}
	\subfloat{\includegraphics[width=0.3\linewidth]{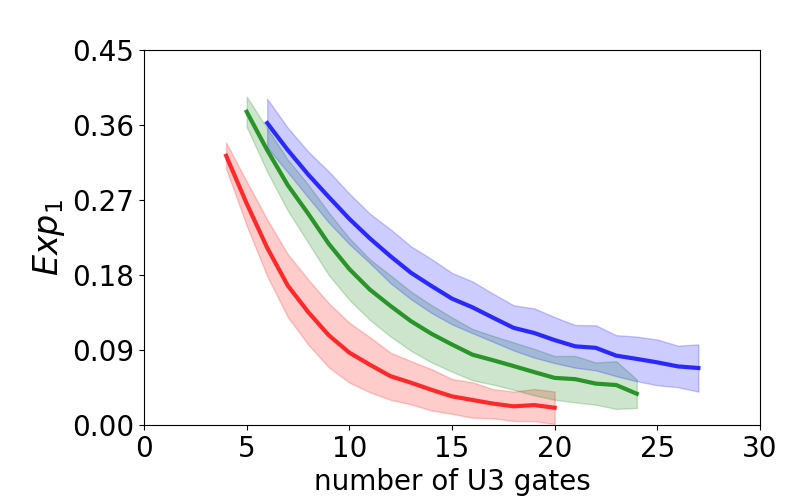} 
		\label{fig:Exp1_U3} }
  \\
	\subfloat{\includegraphics[width=0.3\linewidth]{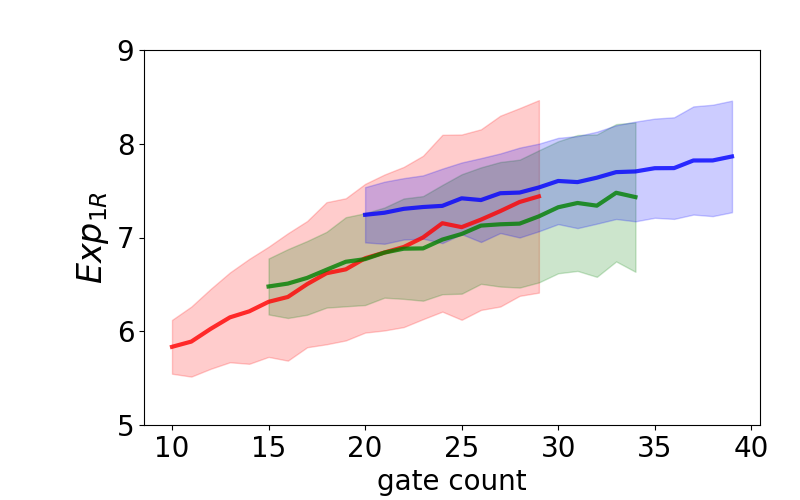}  	\label{fig:Exp2_gateNo} 	}
	\subfloat{\includegraphics[width=0.3\linewidth]{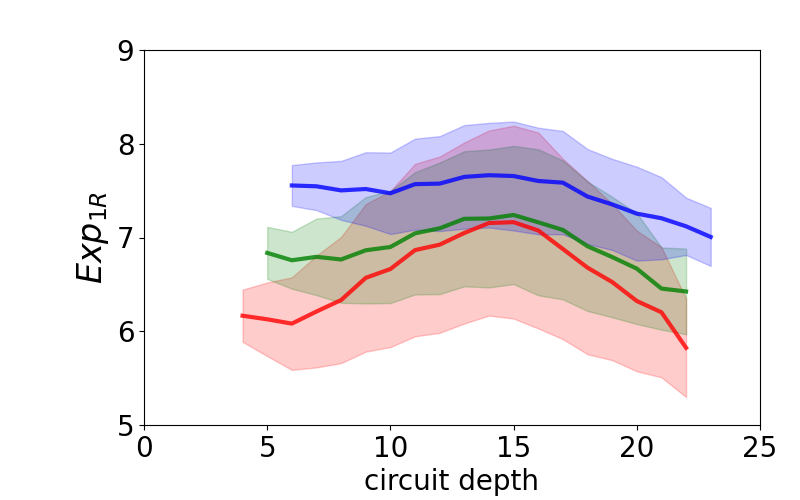} 	\label{fig:Exp2_depth}	}
	\subfloat{\includegraphics[width=0.3\linewidth]{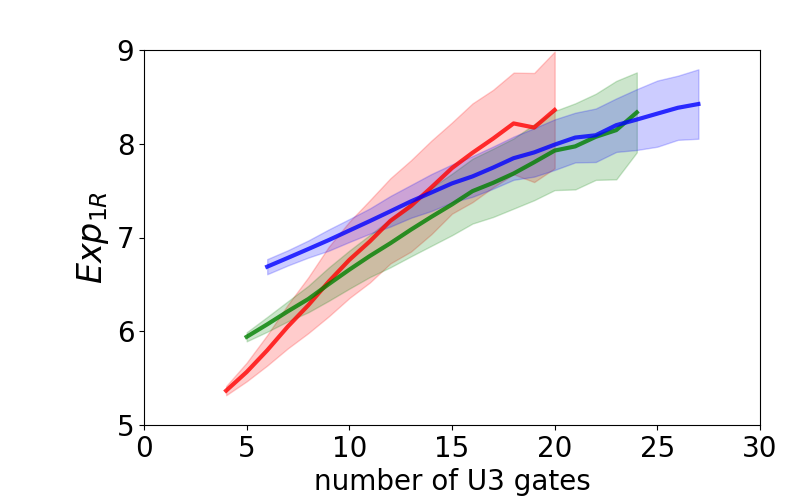} 
		\label{fig:Exp2_U3} }
  \\
	\subfloat{\includegraphics[width=0.3\linewidth]{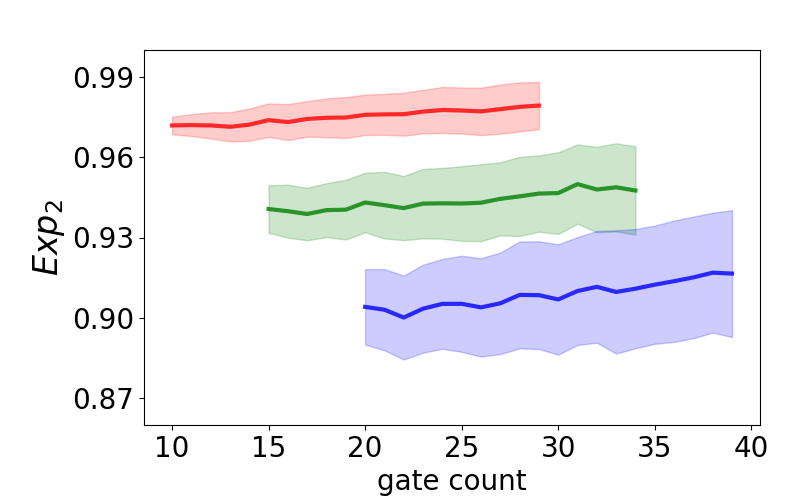}  	\label{fig:Exp3_gateNo} 	}
	\subfloat{\includegraphics[width=0.3\linewidth]{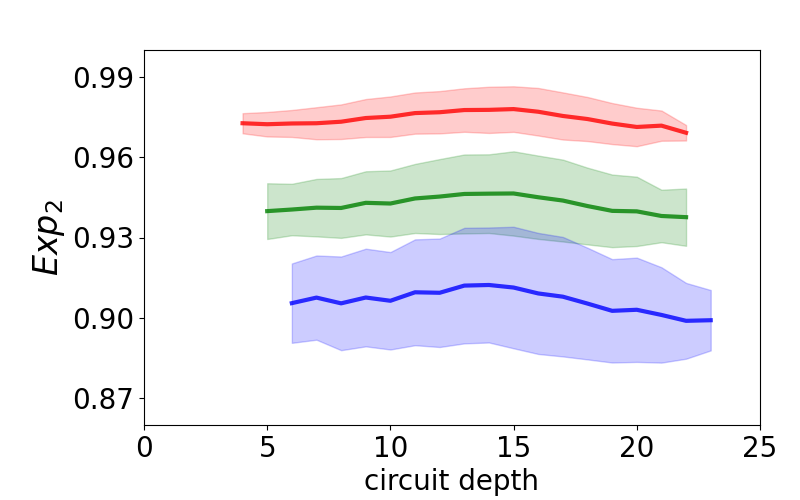} 	\label{fig:Exp3_depth}	}
	\subfloat{\includegraphics[width=0.3\linewidth]{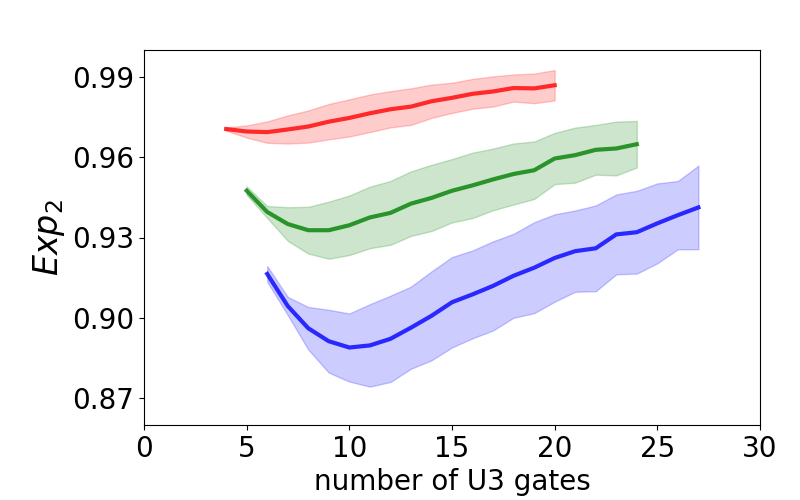} 
		\label{fig:Exp3_U3} }
  \\
	\subfloat{\includegraphics[width=0.3\linewidth]{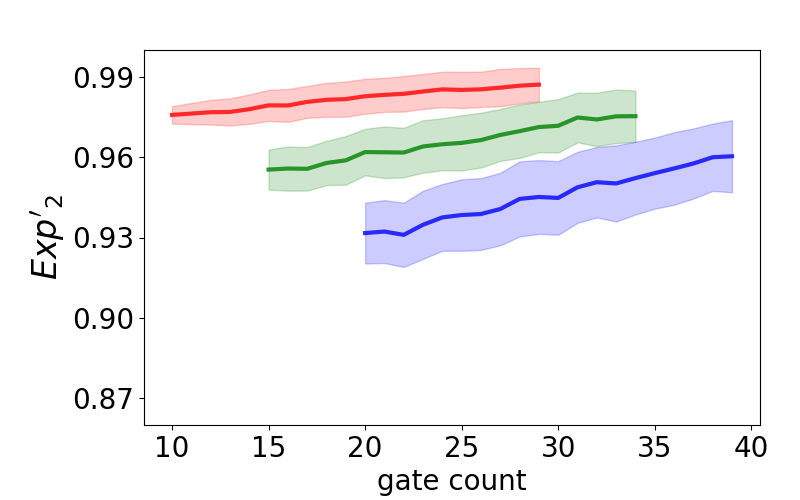}  	\label{fig:Exp4_gateNo} 	}
	\subfloat{\includegraphics[width=0.3\linewidth]{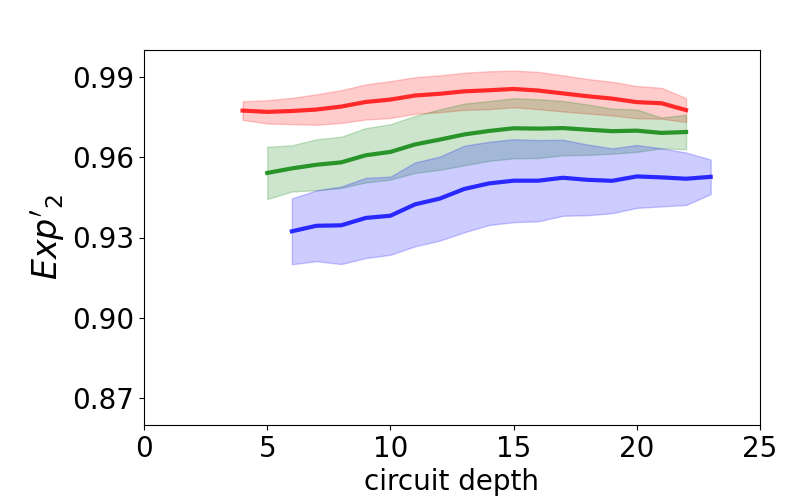} 	\label{fig:Exp4_depth}	}
	\subfloat{\includegraphics[width=0.3\linewidth]{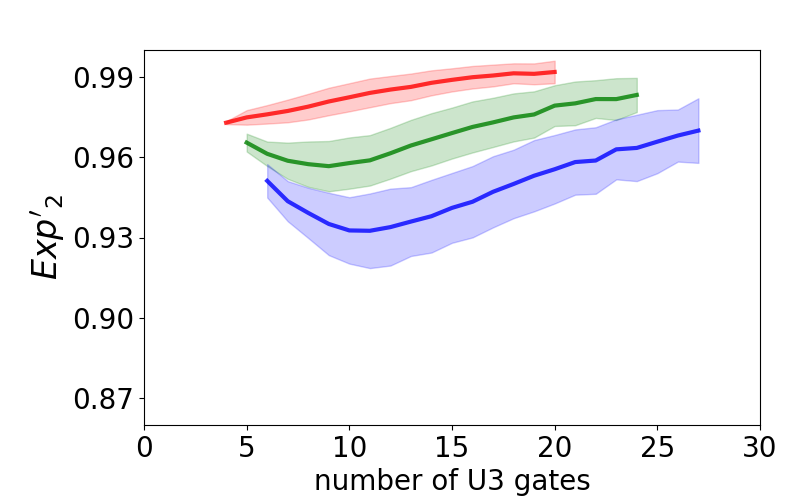} 
		\label{fig:Exp4_U3} }
  \caption{Relationship between various expressibility measures and circuit properties.}
  \label{fig:DatasetProp} 
\end{figure*}

Figure \ref{fig:DatasetProp} illustrates the relationships between expressibility and various circuit properties, including the number of qubits, number of gates, circuit depth and number of U3 gates. $Exp_1$ adopts KL divergence to measure the consistency between the fidelity distribution of the PQC and Harr distribution, where a smaller $Exp_1$ indicates better expressibility. $Exp_{1R}$ evaluates the negative logarithm function of the normalized expressibility, with a higher $Exp_{1R}$ indicating better expressibility. $Exp_2$ calculates the MMD distance between PQC outputs and uniform points in the simplex, where a higher $Exp_2$ denotes better expressibility. 

The leftmost column of plots in Fig. \ref{fig:DatasetProp} demonstrates that for a given qubit count, increasing the number of quantum gates enhances the circuit's expressibility. This implies that circuits with more gates have greater potential for better expressibility. The middle column of plots shows that  expressibility initially increases  with greater depth but then decreases, indicating that a deeper circuit does not necessarily result in better expressibility. The rightmost column of plots suggests that circuits with more parameters to optimize have a higher likelihood of achieving better expressibility. 

The presence of circuit noise does not significantly alter the overall trend between expressibility and circuit properties. However, noise does increase the expressibility of the circuit to some extent. In conclusion, regardless of the expressibility measure or the presence of noise, circuits with more parameterized gates exhibit better expressibility.

\section{Transformer}\label{ProMethod}

The transformer model is adopted in this work to extract the relationship between circuit characteristics and expressibility. The reason for choosing the transformer model lies in its ability to handle input sequences of varying lengths, a common characteristic of quantum circuits with differing numbers of qubits and gates. This variability is similar to challenges encountered in natural language processing, making the transformer model well-suited for predicting quantum circuit expressibility.

The dataset generated in the previous section is utilized for the expressibility prediction task. The framework of the proposed expressibility estimation method is shown in Fig. \ref{fig:framework}. Each generated circuit is represented as a directed graph, and the circuits are encoded to obtain the gate matrix and adjacency matrix for each circuit. The gate matrix represents node features based on the gate information, while the adjacency matrix provides positional information of each gate. These matrices collectively form the graph encoding. In the second step, we apply four different expressibility measures to calculate the expressibility of each PQC in the dataset, yielding four distinct labels for each circuit sample. In the third step, the predictor, which includes graph embedding, transformer layers, and fully connected layers, is trained to predict the expressibility of the circuits. 

\begin{figure*}[!tbp]
\centering
\includegraphics[width=0.95\linewidth]{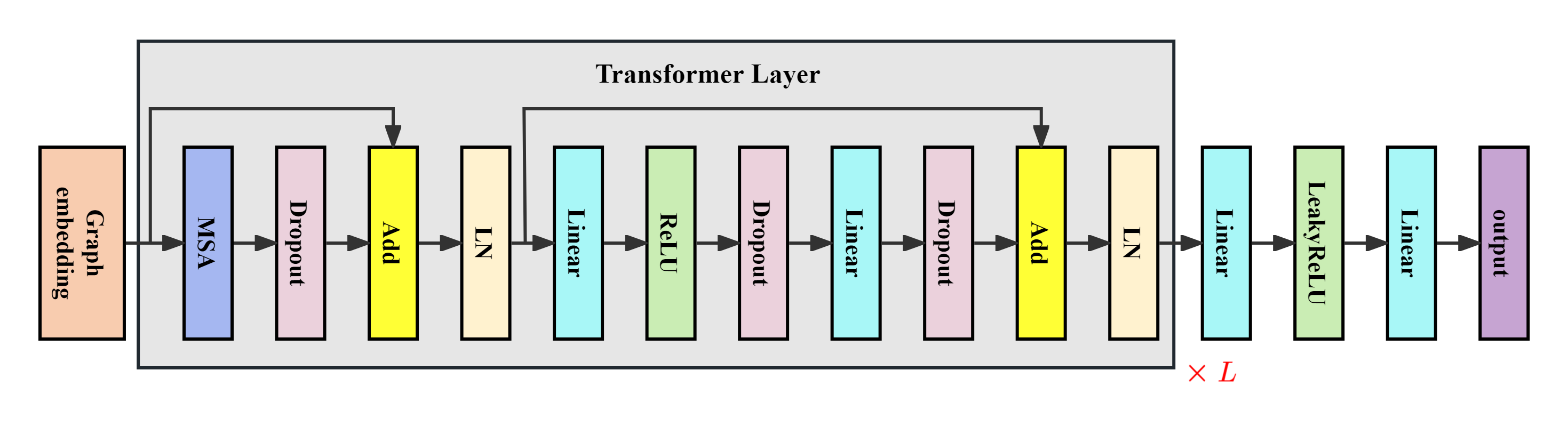} 
\caption{Transformer architecture.}
\label{fig:transformer}
\end{figure*}

The detailed architecture of the transformer model is shown in Fig. \ref{fig:transformer}. Firstly, both the gate matrix and adjacency matrix are embedded into the same hidden dimensional space. These learned embeddings are then added to represent a circuit. A transformer model with $L$ transformer encoder layers is applied to extract characteristics between circuits and their expressibilities. If $L$ ($L \geqslant 2$) transformer encoder layers are implemented, then the output of the $(L-1)$-th layer serves as the input of the $L$-th layer. A single transformer layer is constructed using a multi-head self-attention (MSA) layer, dropout layer, layer normalization (LN), linear layer and ReLU activation layer. Subsequently, fully connected layers with the LeakyReLU activation function are employed to predict the expressibility based on the output of the transformer layer. The prediction model is updated using mean squared error (MSE) loss between the predicted results and target values.

\section{Results}\label{Experiment}

This section describe the numerical results of our approach. We analyze whether the performance of the transformer model is influenced by factors such as the network structure, the type of expressibility measure used, and the presence of noise in the circuits.

\subsection{Evaluation Metrics}

Four commonly used metrics—root mean square error (RMSE), $R^2$, Spearman correlation coefficient ($\rho$), and Kendall correlation coefficient ($\tau$)—are adopted to evaluate the accuracy, robustness and reliability of the trained model. In the following definitions, $\hat{y}_i$ stands for the predicted expressibility of the $i$-th circuit, $y_i$ is the ground-truth expressibility of the $i$-th circuit,  $n$ is the number of quantum circuits, $\bar{y}$ is the mean of circuits' ground-truth expressibility,  $r(y_i)$ is the expressibility rank of the $i$-th circuit, $\text{sgn}$ is the sign function.

(1) RMSE measures the average deviation between predicted expressibilities and ground-truth expressibilities. It is defined as:
\begin{equation}\label{equ:rmse}
	\text{RMSE} = \sqrt{\frac{1}{n} \sum_{i=1}^{n} (\hat{y}_i - y_i)^2}.
\end{equation}
A lower RMSE signifies better prediction accuracy.


(2) $R^2$ is used to estimate the reliability of model fitting results. It is defined as:
\begin{equation}\label{equ:r2}
	R^2 = 1 - \frac{\sum_{i=1}^{n} (y_i - \hat{y}_i)^2}{\sum_{i=1}^{n} (y_i - \bar{y})^2}.
\end{equation}
Higher $R^2$ values (close to 1) indicate better model performance.

(3) The Spearman correlation coefficient ($\rho$) assesses the monotonic ranking relationship between predicted and ground-truth expressibilities. It is defined as:
 \begin{equation}\label{equ:Spearman}
 	\rho = 1 - \frac{6 \sum_{i=1}^{n} (r(y_i)-r( \hat{y}_i))^2}{n(n^2 - 1)}.
 \end{equation}
 Higher $\rho$ values (close to 1)  indicate that the rankings of expressibilities are well-preserved in the predictions. 
 
(4) The Kendall correlation coefficient ($\tau$) measures the ordinal association between predicted and ground-truth expressibilities. It is defined as:
\begin{equation}\label{equ:Kendall}
	\tau =\frac{2}{n(n- 1)} \sum_{i<j} \mathrm{sgn}(y_i-y_j)\mathrm{sgn}(\hat{y}_i-\hat{y}_j).
\end{equation}
Higher $\tau$ values indicate a strong ordinal relationship between the expressibility ranking.

\subsection{Settings}\label{Expersetting}

The dataset is randomly divided into a training set (80\%) and a testing set (20\%). During training, the mean square error (MSE) between the ground-truth expressibility and the model's predicted value is used as the loss function. The Adam optimizer is employed with a learning rate of 0.001, and the CosineAnnealingLR scheduler is used to progressively reduce the learning rate. The transformer model is trained with a batch size of 64 over 100 epochs.

To comprehensively compare the efficacy of the transformer model in estimating four different expressibility measures, we trained four separate transformer models using the entire set of circuits in the training set. Each model undergoes ten independent training runs to ensure the reliability of the results.

We analyze the influence of various transformer structures on model performance by investigating the RMSE result with different number of heads, hidden layers and dimensions of hidden layers. For instance, in Fig. \ref{fig:Exp1-1}, ``1-1-16'' on the x-axis represents a transformer model with 1 head, 1 hidden layer, and a hidden dimension of 16. The pink box represents the interquartile range (IQR), showing the 25th and 75th percentiles of the RMSE values for each trained model, while the green horizontal line indicates the median RMSE value. The lower and upper whiskers represent the minimum and maximum RMSE values within 1.5 times the IQR from the quartiles, respectively.

Therefore, the closer the green horizontal line is to the x-axis, the smaller the loss of the trained model. Additionally, a lower height of the pink box indicates a smaller difference between the 25th and 75th percentiles, signifying stability in performance and greater robustness of the trained model using this structure. Consequently, a structure with a lower median (green line) and a shorter pink box height indicates better prediction performance.

\subsection{Results of $Exp_1$ Estimation}

\begin{figure*}[!htbp]
	\centering
	\includegraphics[width=0.75\linewidth]{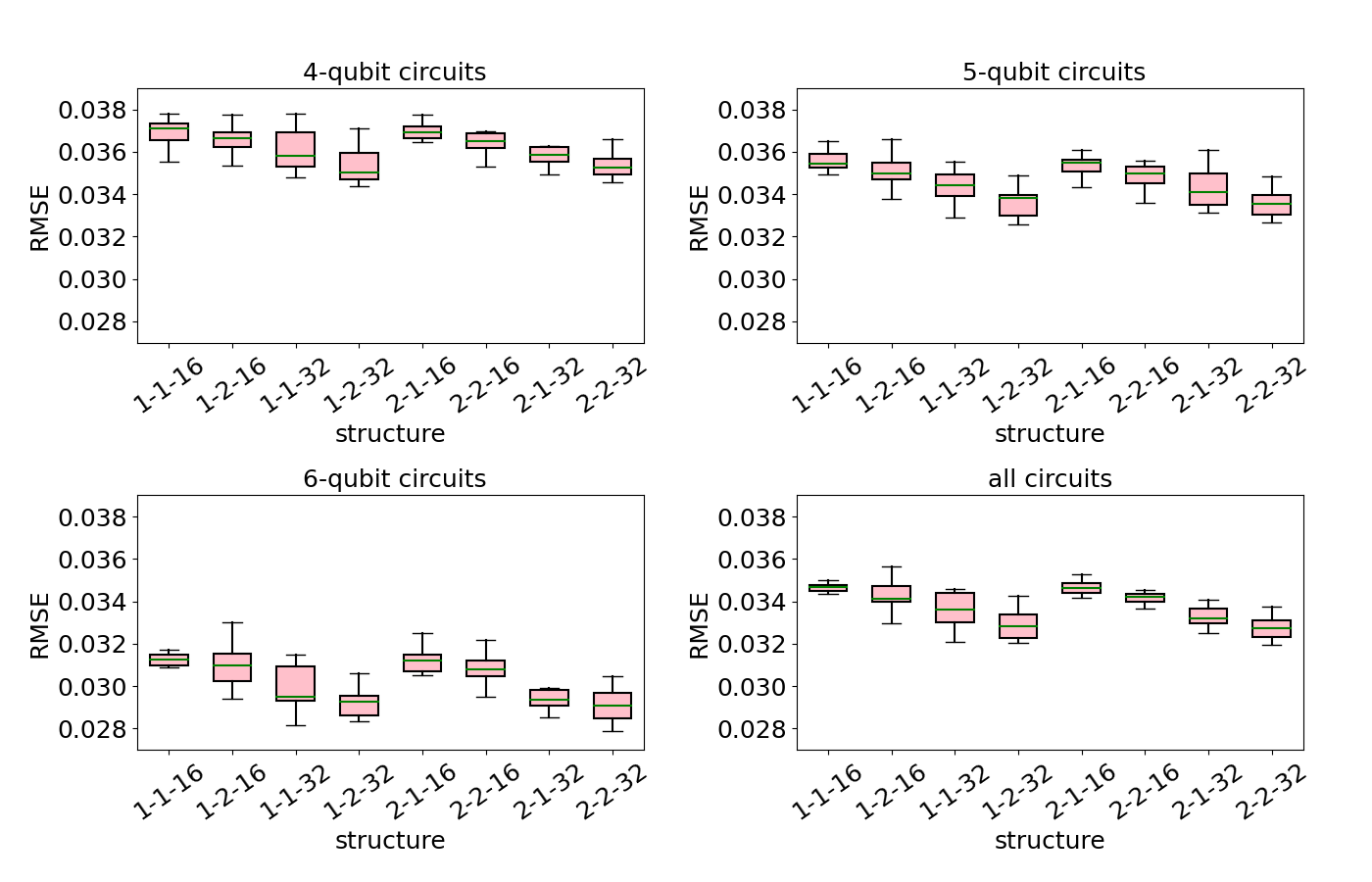}
	\caption{The RMSE of $Exp_1$ prediction across various transformer structures.}
	\label{fig:Exp1-1}
\end{figure*}

As shown in Fig. \ref{fig:Exp1-1}, different network structures have varying impacts on the RMSE. When predicting $Exp_1$ for 4-qubit circuits, the RMSE varies across structures, with an average of approximately 0.036. The ``1-2-32'' structure achieves the lowest RMSE, while the ``2-1-16'' structure has the highest. For 5-qubit circuits, the ``1-2-32'' and ``2-2-32'' structures achieve low median RMSE values. For 6-qubit circuits, the ``1-2-32'' and ``2-1-32'' structures obtain low median RMSE. When predicting $Exp_1$ across all circuits, the ``1-2-32,'' ``2-1-32,'' and ``2-2-32'' structures perform well. Based on these results, we select the ``1-2-32'' structure as an optimal trade-off.

\begin{figure*}[!htbp]
    \centering
    \includegraphics[width=0.75\linewidth]{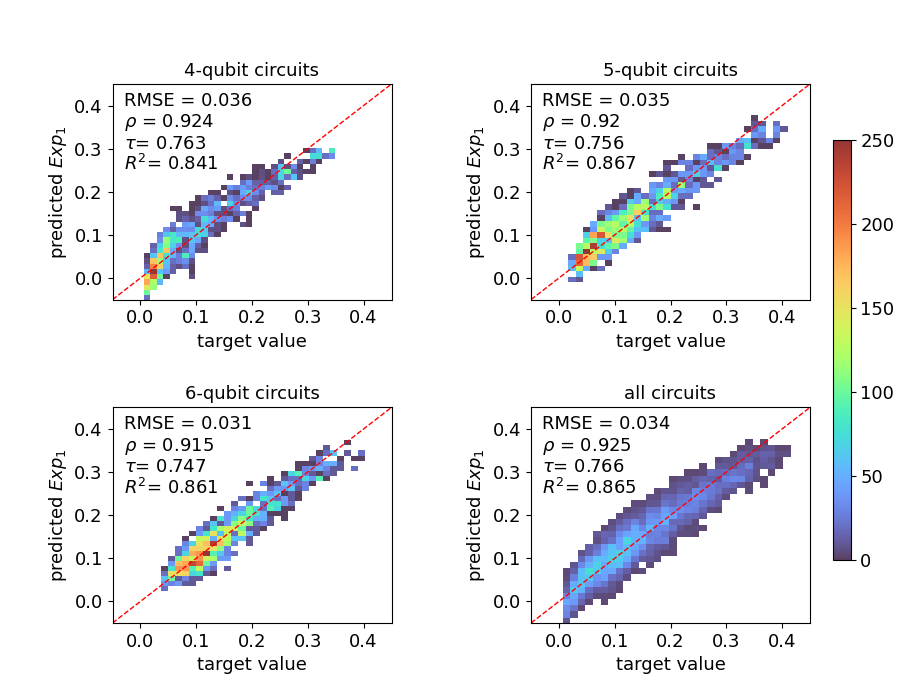}
    \caption{Scatter plots of the relationship between the predicted and ground-truth $Exp_1$.}
    \label{fig:Exp1-2}
\end{figure*}

Figure \ref{fig:Exp1-2} displays scatter plots comparing the predicted $Exp_1$ values with the ground-truth values. In the plots, dark red regions indicate a higher concentration of circuits, and proximity to the red slanted dotted line reflects better prediction performance. The RMSE of the prediction is below 0.036, demonstrating that the model's predictions are very close to the actual expressibility values and indicating high performance. Additionally, the Spearman correlation coefficient ($\rho$) exceeds 0.915, showing that the rankings of expressibility are well-preserved. The Kendall correlation coefficient ($\tau$) is larger than 0.747, confirming a strong association between predicted and actual expressibility values. The $R^2$ value exceeds 0.841, indicating that the model's fit is reliable.

The results demonstrate that the trained transformer model can reliably predict the $Exp_1$ expressibility of noiseless PQCs. Its accurate predictions of expressibility enable the efficient identification of expressive PQC architectures, reducing the need for exhaustive parameter optimization and thereby conserving both time and computational resources.

\subsection{Results of $Exp_{1R}$ Estimation}

\begin{figure*}[!htbp]
    \centering
    \includegraphics[width=0.75\linewidth]{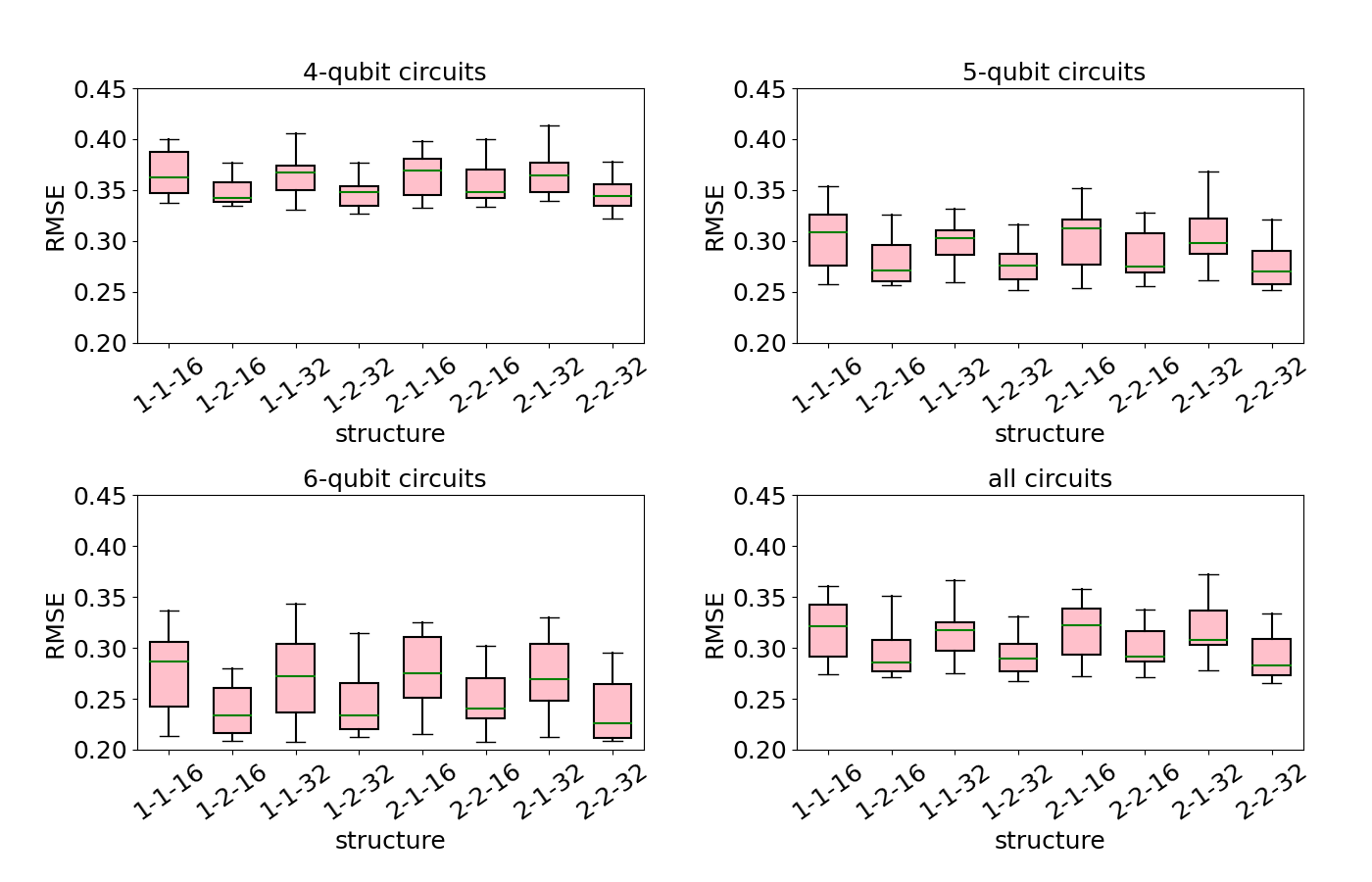}
    \caption{The RMSE of $Exp_{1R}$ prediction across various transformer structures.}
    \label{fig:Exp1-3}
\end{figure*}

The $Exp_{1R}$ measure is calculated as the negative logarithm of $Exp_1$ compared to the expressibility of an idle circuit $Exp_1(\text{Idle})$. The ground-truth values of $Exp_{1R}$ range from 5 to 9, while those of $Exp_1$ ranges from 0 to 0.5. Consequently, the RMSE of the trained model using $Exp_{1R}$ is larger than that using $Exp_{1}$. Figure \ref{fig:Exp1-3} illustrates that different transformer model structures yield significantly varying prediction errors.  
Overall, the ``2-2-32'' structure achieves lower RMSE and demonstrates robust results across four prediction tasks. Therefore, the ``2-2-32'' structure is used for $Exp_{1R}$ prediction.

\begin{figure*}[!htbp]
    \centering
    \includegraphics[width=0.75\linewidth]{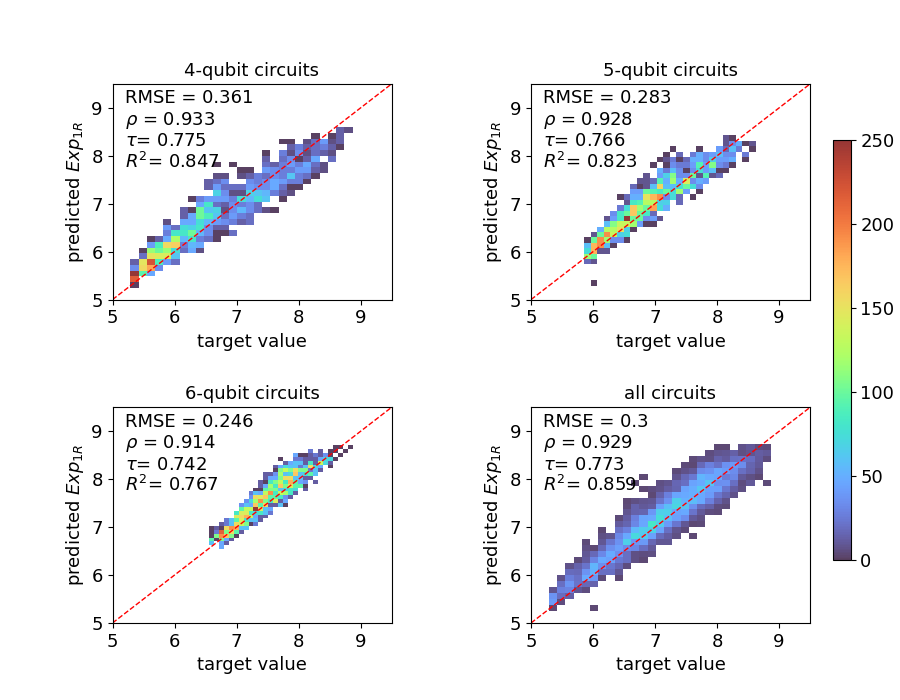}
    \caption{Scatter plots of the relationship between the predicted and ground-truth $Exp_{1R}$.}
    \label{fig:Exp1-4}
\end{figure*}

In Fig. \ref{fig:Exp1-4}, we observe results similar to those in Fig. \ref{fig:Exp1-2}. This consistency indicates that the trained model is accurate and generalizable across different types of expressibility measures.

\subsection{Results of $Exp_2$ Estimation}

\begin{figure*}[!htbp]
    \centering
    \includegraphics[width=0.75\linewidth]{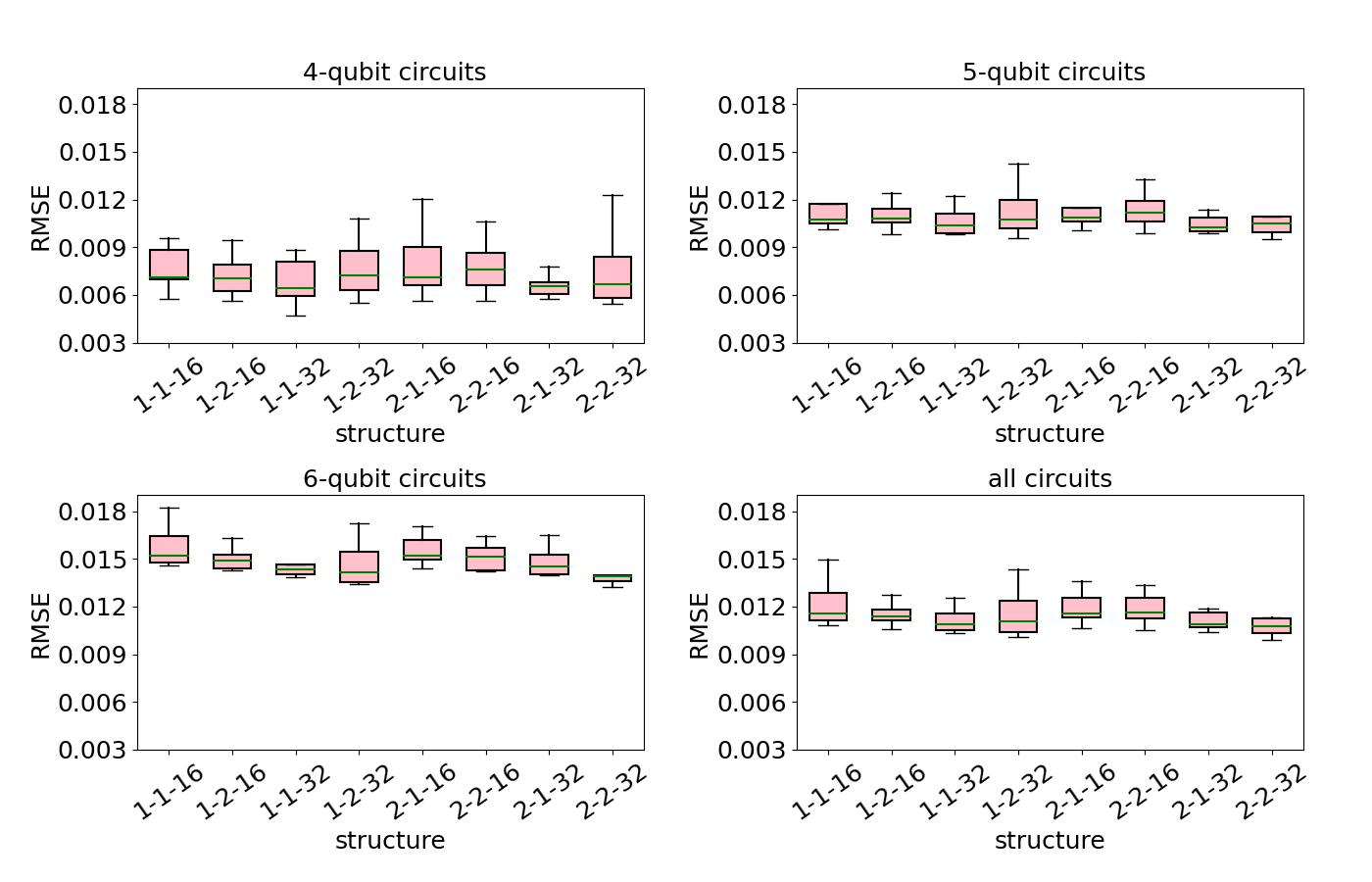}
    \caption{The RMSE of $Exp_{2}$ prediction across various transformer structures.}
    \label{fig:Exp2-1}
\end{figure*}

In the definition of $Exp_2$, the MMD is adopted to evaluate the discrepancy between the PQC outputs and uniform outputs in the simplex. This approach eliminates the need to calculate the fidelity of the PQC-generated states, thereby reducing computational cost.  From Fig. \ref{fig:Exp2-1}, we observe that different network structures have little impact on the predictive performance of the model, as the median RMSE values of these structures are approximately the same.
Some structures, such as ``1-2-16'' and ``2-1-32,''  yield more stable results, while others, like ``1-2-32'' and ``2-2-32,'' are less stable. Overall, the ``2-1-32'' structure emerges as the best choice for predicting  $Exp_2$.

\begin{figure*}[!htbp]
    \centering
    \includegraphics[width=0.75\linewidth]{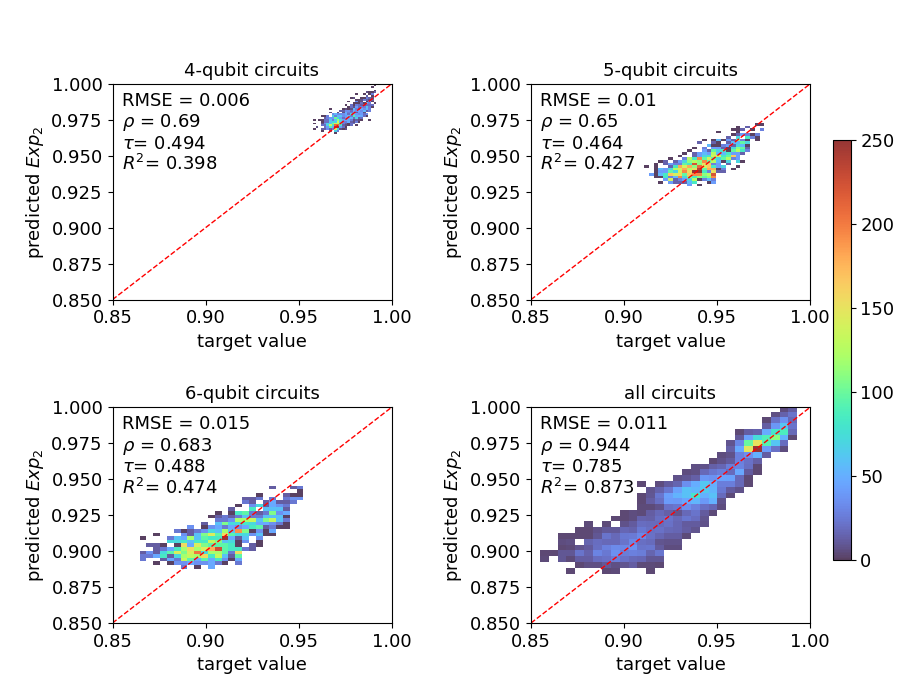}
    \caption{Scatter plots of the relationship between the predicted and ground-truth $Exp_{2}$.}
    \label{fig:Exp2-2}
\end{figure*}

From Fig. \ref{fig:Exp2-2}, it is evident that circuits with 4 qubits have larger expressibility values than those with 6 qubits, and the RMSE of the trained model on 4-qubit circuits is the smallest compared to the other three cases. However, the fitting results of the trained model on 4-qubit circuits are worse than on circuits with other qubit counts, as indicated by the smaller $R^2$ value. This is primarily because the range of expressibility values among 4-qubit circuits is smaller than other cases. In contrast, for all circuits, a larger expressibility range achives a larger $R^2$ value. This indicates that the transformer model has a strong ability to learn and generalize the relationship between circuit architecture and expressibility.

\subsection{Results of $Exp'_2$ Estimation}

\begin{figure*}[!htbp]
    \centering
    \includegraphics[width=0.75\linewidth]{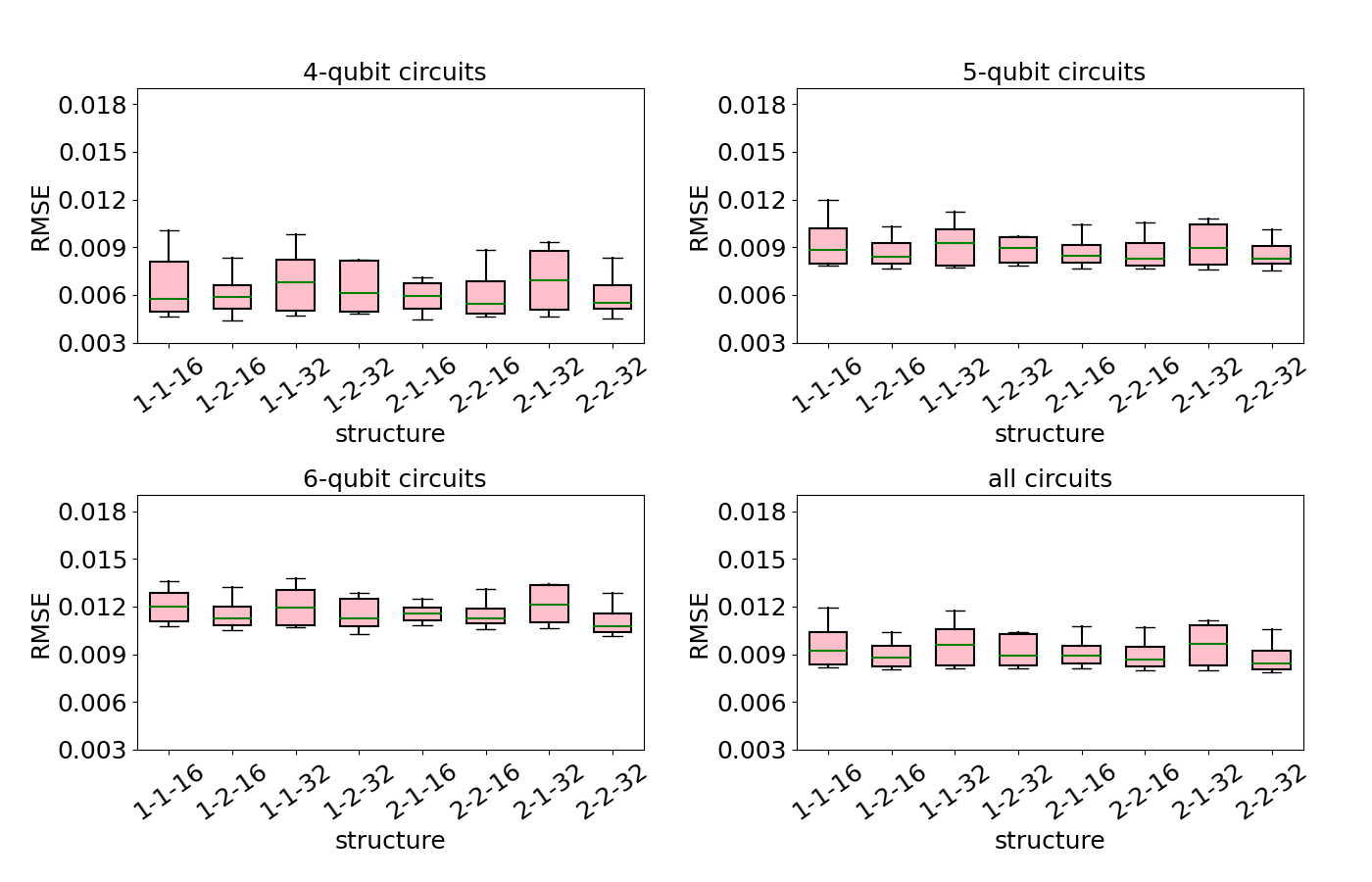}
    \caption{The RMSE of $Exp'_2$ prediction across various transformer structures.}
    \label{fig:Exp2-3}
\end{figure*}

\begin{figure*}[!htbp]
    \centering
    \includegraphics[width=0.75\linewidth]{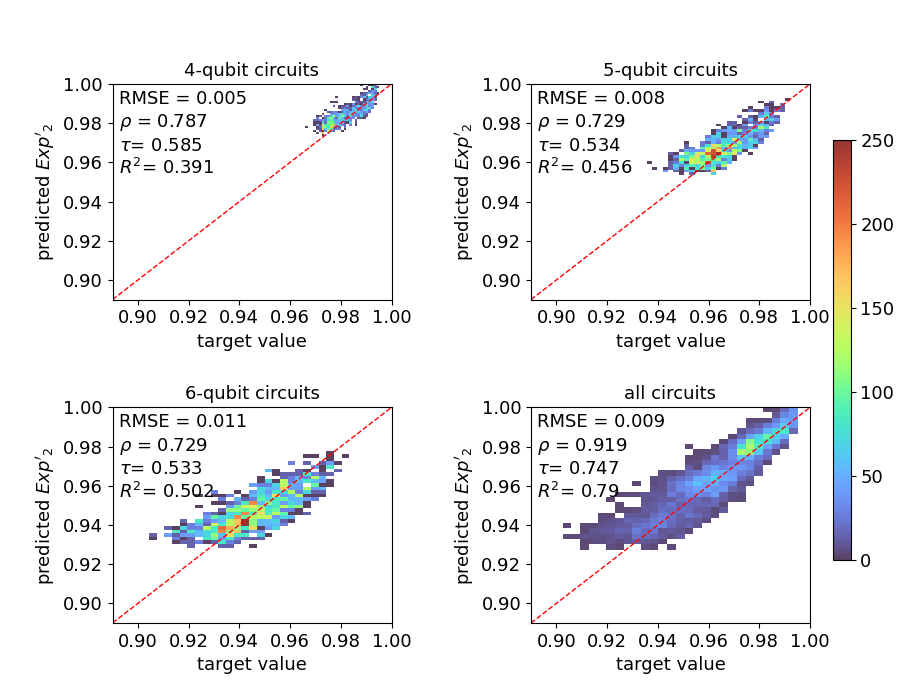}
    \caption{Scatter plots of the relationship between the predicted and ground-truth $Exp'_{2}$.}
    \label{fig:Exp2-4}
\end{figure*}

From Fig. \ref{fig:Exp2-3} , we observe results similar to those in Fig. \ref{fig:Exp2-1}, confirming that the ``1-2-16'' structure is appropriate for this expressibility measure. Comparing  Fig.  \ref{fig:Exp2-4} with Fig. \ref{fig:Exp2-2}, it is evident that the model trained on noisy circuits performs better than the model trained on noiseless circuits. This suggests that the presence of noise enhances the distinguishing characteristics of the circuits, increasing the likelihood that the trained model can effectively extract differences among the noisy circuits.

In Fig. \ref{fig:Exp2-4}, the model trained on all noisy circuits accurately predicts the expressibility of circuits with an expressibility value greater than 0.97, as indicated by
the bright green region close to the red-slanted dotted line. However, the prediction accuracy decreases for circuits with lower expressibility, around 0.92, although there are fewer circuits in this range. Despite some incorrect predictions, our primary focus is on the more expressive circuits. The trained model successfully predicts these highly expressive circuits with a higher probability, demonstrating its effectiveness.

\section{Conclusion}\label{Conclusion}

The transformer is effective at handling problems with variable-length input sequences. In this work, we construct a dataset containing PQCs that vary in number of qubits and gates. Four expressibility measures are calculated for each PQC in the dataset. After converting PQCs into graphs, we can leverage the transformer's ability to capture the complex relationships between circuit structure and  expressibility.  Numerical results demonstrate the effectiveness of using transformer models in predicting expressibility across various measures of expressibility. 


Future work could explore further enhancements to the model, such as incorporating additional features or exploring ensemble methods to improve prediction performance. Additionally, applying the model to real-world quantum computing tasks would validate its practical utility. Extending the approach to other quantum properties beyond expressibility would also provide a more comprehensive toolkit for quantum circuit design and optimization.

\begin{acknowledgments}
This work is   supported  by  the  Henan  Provincial   Science   and  Technology  Research  Project (No. 222102210258),  Guangdong Basic and Applied Basic Research Foundation (Nos. 2022A1515140116, 2021A1515011985), Jihua Laboratory Scientific Project (No. X210101UZ210), and Innovation Program for Quantum Science and Technology (No. 2021ZD0302901).
\end{acknowledgments}


\begin{thebibliography}{24}%
\makeatletter
\providecommand \@ifxundefined [1]{%
 \@ifx{#1\undefined}
}%
\providecommand \@ifnum [1]{%
 \ifnum #1\expandafter \@firstoftwo
 \else \expandafter \@secondoftwo
 \fi
}%
\providecommand \@ifx [1]{%
 \ifx #1\expandafter \@firstoftwo
 \else \expandafter \@secondoftwo
 \fi
}%
\providecommand \natexlab [1]{#1}%
\providecommand \enquote  [1]{``#1''}%
\providecommand \bibnamefont  [1]{#1}%
\providecommand \bibfnamefont [1]{#1}%
\providecommand \citenamefont [1]{#1}%
\providecommand \href@noop [0]{\@secondoftwo}%
\providecommand \href [0]{\begingroup \@sanitize@url \@href}%
\providecommand \@href[1]{\@@startlink{#1}\@@href}%
\providecommand \@@href[1]{\endgroup#1\@@endlink}%
\providecommand \@sanitize@url [0]{\catcode `\\12\catcode `\$12\catcode
  `\&12\catcode `\#12\catcode `\^12\catcode `\_12\catcode `\%12\relax}%
\providecommand \@@startlink[1]{}%
\providecommand \@@endlink[0]{}%
\providecommand \url  [0]{\begingroup\@sanitize@url \@url }%
\providecommand \@url [1]{\endgroup\@href {#1}{\urlprefix }}%
\providecommand \urlprefix  [0]{URL }%
\providecommand \Eprint [0]{\href }%
\providecommand \doibase [0]{https://doi.org/}%
\providecommand \selectlanguage [0]{\@gobble}%
\providecommand \bibinfo  [0]{\@secondoftwo}%
\providecommand \bibfield  [0]{\@secondoftwo}%
\providecommand \translation [1]{[#1]}%
\providecommand \BibitemOpen [0]{}%
\providecommand \bibitemStop [0]{}%
\providecommand \bibitemNoStop [0]{.\EOS\space}%
\providecommand \EOS [0]{\spacefactor3000\relax}%
\providecommand \BibitemShut  [1]{\csname bibitem#1\endcsname}%
\let\auto@bib@innerbib\@empty
\bibitem [{\citenamefont {Ramezani}\ \emph {et~al.}(2020)\citenamefont
  {Ramezani}, \citenamefont {Sommers}, \citenamefont {Manchukonda},
  \citenamefont {Rahimi},\ and\ \citenamefont {Amirlatifi}}]{mlqc2020}%
  \BibitemOpen
  \bibfield  {author} {\bibinfo {author} {\bibfnamefont {S.~B.}\ \bibnamefont
  {Ramezani}}, \bibinfo {author} {\bibfnamefont {A.}~\bibnamefont {Sommers}},
  \bibinfo {author} {\bibfnamefont {H.~K.}\ \bibnamefont {Manchukonda}},
  \bibinfo {author} {\bibfnamefont {S.}~\bibnamefont {Rahimi}},\ and\ \bibinfo
  {author} {\bibfnamefont {A.}~\bibnamefont {Amirlatifi}},\ }\bibfield  {title}
  {\bibinfo {title} {Machine learning algorithms in quantum computing: A
  survey},\ }in\ \href@noop {} {\emph {\bibinfo {booktitle} {2020 International
  Joint Conference on Neural Networks (IJCNN)}}}\ (\bibinfo {year} {2020})\
  pp.\ \bibinfo {pages} {1--8}\BibitemShut {NoStop}%
\bibitem [{\citenamefont {Mauranyapin}\ \emph {et~al.}(2022)\citenamefont
  {Mauranyapin}, \citenamefont {Terrasson},\ and\ \citenamefont
  {Bowen}}]{QuBiotechnology2022}%
  \BibitemOpen
  \bibfield  {author} {\bibinfo {author} {\bibfnamefont {N.~P.}\ \bibnamefont
  {Mauranyapin}}, \bibinfo {author} {\bibfnamefont {A.}~\bibnamefont
  {Terrasson}},\ and\ \bibinfo {author} {\bibfnamefont {W.~P.}\ \bibnamefont
  {Bowen}},\ }\bibfield  {title} {\bibinfo {title} {Quantum biotechnology},\
  }\href@noop {} {\bibfield  {journal} {\bibinfo  {journal} {Advanced Quantum
  Technologies}\ }\textbf {\bibinfo {volume} {5}},\ \bibinfo {pages} {2100139}
  (\bibinfo {year} {2022})}\BibitemShut {NoStop}%
\bibitem [{\citenamefont {Qi}\ \emph {et~al.}(2024)\citenamefont {Qi},
  \citenamefont {Xiao}, \citenamefont {Liu}, \citenamefont {Gong},\ and\
  \citenamefont {Gani}}]{qi2024variational}%
  \BibitemOpen
  \bibfield  {author} {\bibinfo {author} {\bibfnamefont {H.}~\bibnamefont
  {Qi}}, \bibinfo {author} {\bibfnamefont {S.}~\bibnamefont {Xiao}}, \bibinfo
  {author} {\bibfnamefont {Z.}~\bibnamefont {Liu}}, \bibinfo {author}
  {\bibfnamefont {C.}~\bibnamefont {Gong}},\ and\ \bibinfo {author}
  {\bibfnamefont {A.}~\bibnamefont {Gani}},\ }\bibfield  {title} {\bibinfo
  {title} {Variational quantum algorithms: fundamental concepts, applications
  and challenges},\ }\href@noop {} {\bibfield  {journal} {\bibinfo  {journal}
  {Quantum Information Processing}\ }\textbf {\bibinfo {volume} {23}},\
  \bibinfo {pages} {224} (\bibinfo {year} {2024})}\BibitemShut {NoStop}%
\bibitem [{\citenamefont {Cerezo}\ \emph {et~al.}(2021)\citenamefont {Cerezo},
  \citenamefont {Arrasmith}, \citenamefont {Babbush}, \citenamefont {Benjamin},
  \citenamefont {Endo}, \citenamefont {Fujii}, \citenamefont {McClean},
  \citenamefont {Mitarai}, \citenamefont {Yuan}, \citenamefont {Cincio} \emph
  {et~al.}}]{cerezo2021variational}%
  \BibitemOpen
  \bibfield  {author} {\bibinfo {author} {\bibfnamefont {M.}~\bibnamefont
  {Cerezo}}, \bibinfo {author} {\bibfnamefont {A.}~\bibnamefont {Arrasmith}},
  \bibinfo {author} {\bibfnamefont {R.}~\bibnamefont {Babbush}}, \bibinfo
  {author} {\bibfnamefont {S.~C.}\ \bibnamefont {Benjamin}}, \bibinfo {author}
  {\bibfnamefont {S.}~\bibnamefont {Endo}}, \bibinfo {author} {\bibfnamefont
  {K.}~\bibnamefont {Fujii}}, \bibinfo {author} {\bibfnamefont {J.~R.}\
  \bibnamefont {McClean}}, \bibinfo {author} {\bibfnamefont {K.}~\bibnamefont
  {Mitarai}}, \bibinfo {author} {\bibfnamefont {X.}~\bibnamefont {Yuan}},
  \bibinfo {author} {\bibfnamefont {L.}~\bibnamefont {Cincio}}, \emph
  {et~al.},\ }\bibfield  {title} {\bibinfo {title} {Variational quantum
  algorithms},\ }\href@noop {} {\bibfield  {journal} {\bibinfo  {journal}
  {Nature Reviews Physics}\ }\textbf {\bibinfo {volume} {3}},\ \bibinfo {pages}
  {625} (\bibinfo {year} {2021})}\BibitemShut {NoStop}%
\bibitem [{\citenamefont {Gong}\ \emph {et~al.}(2024)\citenamefont {Gong},
  \citenamefont {Pei}, \citenamefont {Zhang},\ and\ \citenamefont
  {Zhou}}]{gong2024quantum}%
  \BibitemOpen
  \bibfield  {author} {\bibinfo {author} {\bibfnamefont {L.-H.}\ \bibnamefont
  {Gong}}, \bibinfo {author} {\bibfnamefont {J.-J.}\ \bibnamefont {Pei}},
  \bibinfo {author} {\bibfnamefont {T.-F.}\ \bibnamefont {Zhang}},\ and\
  \bibinfo {author} {\bibfnamefont {N.-R.}\ \bibnamefont {Zhou}},\ }\bibfield
  {title} {\bibinfo {title} {Quantum convolutional neural network based on
  variational quantum circuits},\ }\href@noop {} {\bibfield  {journal}
  {\bibinfo  {journal} {Optics Communications}\ }\textbf {\bibinfo {volume}
  {550}},\ \bibinfo {pages} {129993} (\bibinfo {year} {2024})}\BibitemShut
  {NoStop}%
\bibitem [{\citenamefont {Fang}\ \emph {et~al.}(2024)\citenamefont {Fang},
  \citenamefont {Zhang},\ and\ \citenamefont {Situ}}]{Situ2024}%
  \BibitemOpen
  \bibfield  {author} {\bibinfo {author} {\bibfnamefont {P.}~\bibnamefont
  {Fang}}, \bibinfo {author} {\bibfnamefont {C.}~\bibnamefont {Zhang}},\ and\
  \bibinfo {author} {\bibfnamefont {H.}~\bibnamefont {Situ}},\ }\bibfield
  {title} {\bibinfo {title} {Quantum state clustering algorithm based on
  variational quantum circuit},\ }\href@noop {} {\bibfield  {journal} {\bibinfo
   {journal} {Quantum Information Processing}\ }\textbf {\bibinfo {volume}
  {23}},\ \bibinfo {pages} {125} (\bibinfo {year} {2024})}\BibitemShut
  {NoStop}%
\bibitem [{\citenamefont {He}\ \emph {et~al.}(2024)\citenamefont {He},
  \citenamefont {Deng}, \citenamefont {Zheng}, \citenamefont {Li},\ and\
  \citenamefont {Situ}}]{zhimin2024aaai}%
  \BibitemOpen
  \bibfield  {author} {\bibinfo {author} {\bibfnamefont {Z.}~\bibnamefont
  {He}}, \bibinfo {author} {\bibfnamefont {M.}~\bibnamefont {Deng}}, \bibinfo
  {author} {\bibfnamefont {S.}~\bibnamefont {Zheng}}, \bibinfo {author}
  {\bibfnamefont {L.}~\bibnamefont {Li}},\ and\ \bibinfo {author}
  {\bibfnamefont {H.}~\bibnamefont {Situ}},\ }\bibfield  {title} {\bibinfo
  {title} {Training-free quantum architecture search},\ }in\ \href@noop {}
  {\emph {\bibinfo {booktitle} {38th AAAI Conference on Artificial Intelligence
  (AAAI)}}}\ (\bibinfo {year} {2024})\BibitemShut {NoStop}%
\bibitem [{\citenamefont {Zhang}\ \emph {et~al.}(2022)\citenamefont {Zhang},
  \citenamefont {Hsieh}, \citenamefont {Zhang},\ and\ \citenamefont
  {Yao}}]{zhang2022differentiable}%
  \BibitemOpen
  \bibfield  {author} {\bibinfo {author} {\bibfnamefont {S.-X.}\ \bibnamefont
  {Zhang}}, \bibinfo {author} {\bibfnamefont {C.-Y.}\ \bibnamefont {Hsieh}},
  \bibinfo {author} {\bibfnamefont {S.}~\bibnamefont {Zhang}},\ and\ \bibinfo
  {author} {\bibfnamefont {H.}~\bibnamefont {Yao}},\ }\bibfield  {title}
  {\bibinfo {title} {Differentiable quantum architecture search},\ }\href@noop
  {} {\bibfield  {journal} {\bibinfo  {journal} {Quantum Science and
  Technology}\ }\textbf {\bibinfo {volume} {7}},\ \bibinfo {pages} {045023}
  (\bibinfo {year} {2022})}\BibitemShut {NoStop}%
\bibitem [{\citenamefont {Zhang}\ \emph {et~al.}(2021)\citenamefont {Zhang},
  \citenamefont {Hsieh}, \citenamefont {Zhang},\ and\ \citenamefont
  {Yao}}]{zhangHsieh2021}%
  \BibitemOpen
  \bibfield  {author} {\bibinfo {author} {\bibfnamefont {S.-X.}\ \bibnamefont
  {Zhang}}, \bibinfo {author} {\bibfnamefont {C.-Y.}\ \bibnamefont {Hsieh}},
  \bibinfo {author} {\bibfnamefont {S.}~\bibnamefont {Zhang}},\ and\ \bibinfo
  {author} {\bibfnamefont {H.}~\bibnamefont {Yao}},\ }\bibfield  {title}
  {\bibinfo {title} {Neural predictor based quantum architecture search},\
  }\href@noop {} {\bibfield  {journal} {\bibinfo  {journal} {Machine Learning:
  Science and Technology}\ }\textbf {\bibinfo {volume} {2}},\ \bibinfo {pages}
  {045027} (\bibinfo {year} {2021})}\BibitemShut {NoStop}%
\bibitem [{\citenamefont {Situ}\ \emph {et~al.}(2024)\citenamefont {Situ},
  \citenamefont {He}, \citenamefont {Zheng},\ and\ \citenamefont
  {Li}}]{Situ_2024}%
  \BibitemOpen
  \bibfield  {author} {\bibinfo {author} {\bibfnamefont {H.}~\bibnamefont
  {Situ}}, \bibinfo {author} {\bibfnamefont {Z.}~\bibnamefont {He}}, \bibinfo
  {author} {\bibfnamefont {S.}~\bibnamefont {Zheng}},\ and\ \bibinfo {author}
  {\bibfnamefont {L.}~\bibnamefont {Li}},\ }\bibfield  {title} {\bibinfo
  {title} {Distributed quantum architecture search},\ }\href@noop {} {\bibfield
   {journal} {\bibinfo  {journal} {arXiv preprint arXiv:2403.06214, 2024}\ }
  (\bibinfo {year} {2024})}\BibitemShut {NoStop}%
\bibitem [{\citenamefont {He}\ \emph {et~al.}(2021)\citenamefont {He},
  \citenamefont {Li}, \citenamefont {Zheng}, \citenamefont {Li},\ and\
  \citenamefont {Situ}}]{he2021variational}%
  \BibitemOpen
  \bibfield  {author} {\bibinfo {author} {\bibfnamefont {Z.}~\bibnamefont
  {He}}, \bibinfo {author} {\bibfnamefont {L.}~\bibnamefont {Li}}, \bibinfo
  {author} {\bibfnamefont {S.}~\bibnamefont {Zheng}}, \bibinfo {author}
  {\bibfnamefont {Y.}~\bibnamefont {Li}},\ and\ \bibinfo {author}
  {\bibfnamefont {H.}~\bibnamefont {Situ}},\ }\bibfield  {title} {\bibinfo
  {title} {Variational quantum compiling with double {Q}-learning},\
  }\href@noop {} {\bibfield  {journal} {\bibinfo  {journal} {New Journal of
  Physics}\ }\textbf {\bibinfo {volume} {23}},\ \bibinfo {pages} {033002}
  (\bibinfo {year} {2021})}\BibitemShut {NoStop}%
\bibitem [{\citenamefont {Altares-L{\'o}pez}\ \emph {et~al.}(2021)\citenamefont
  {Altares-L{\'o}pez}, \citenamefont {Ribeiro},\ and\ \citenamefont
  {Garc{\'\i}a-Ripoll}}]{altares2021automatic}%
  \BibitemOpen
  \bibfield  {author} {\bibinfo {author} {\bibfnamefont {S.}~\bibnamefont
  {Altares-L{\'o}pez}}, \bibinfo {author} {\bibfnamefont {A.}~\bibnamefont
  {Ribeiro}},\ and\ \bibinfo {author} {\bibfnamefont {J.~J.}\ \bibnamefont
  {Garc{\'\i}a-Ripoll}},\ }\bibfield  {title} {\bibinfo {title} {Automatic
  design of quantum feature maps},\ }\href@noop {} {\bibfield  {journal}
  {\bibinfo  {journal} {Quantum Science and Technology}\ }\textbf {\bibinfo
  {volume} {6}},\ \bibinfo {pages} {045015} (\bibinfo {year}
  {2021})}\BibitemShut {NoStop}%
\bibitem [{\citenamefont {Sim}\ \emph {et~al.}(2019)\citenamefont {Sim},
  \citenamefont {Johnson},\ and\ \citenamefont
  {Aspuru-Guzik}}]{HybridExpressibility2019}%
  \BibitemOpen
  \bibfield  {author} {\bibinfo {author} {\bibfnamefont {S.}~\bibnamefont
  {Sim}}, \bibinfo {author} {\bibfnamefont {P.~D.}\ \bibnamefont {Johnson}},\
  and\ \bibinfo {author} {\bibfnamefont {A.}~\bibnamefont {Aspuru-Guzik}},\
  }\bibfield  {title} {\bibinfo {title} {Expressibility and entangling
  capability of parameterized quantum circuits for hybrid quantum-classical
  algorithms},\ }\href@noop {} {\bibfield  {journal} {\bibinfo  {journal}
  {Advanced Quantum Technologies}\ }\textbf {\bibinfo {volume} {2}},\ \bibinfo
  {pages} {1900070} (\bibinfo {year} {2019})}\BibitemShut {NoStop}%
\bibitem [{\citenamefont {Rasmussen}\ \emph {et~al.}(2020)\citenamefont
  {Rasmussen}, \citenamefont {Loft}, \citenamefont {Bkkegaard}, \citenamefont
  {Kues},\ and\ \citenamefont {Zinner}}]{KL_R2020}%
  \BibitemOpen
  \bibfield  {author} {\bibinfo {author} {\bibfnamefont {S.~E.}\ \bibnamefont
  {Rasmussen}}, \bibinfo {author} {\bibfnamefont {N.~J.~S.}\ \bibnamefont
  {Loft}}, \bibinfo {author} {\bibfnamefont {T.}~\bibnamefont {Bkkegaard}},
  \bibinfo {author} {\bibfnamefont {M.}~\bibnamefont {Kues}},\ and\ \bibinfo
  {author} {\bibfnamefont {N.~T.}\ \bibnamefont {Zinner}},\ }\bibfield  {title}
  {\bibinfo {title} {Reducing the amount of single‐qubit rotations in vqe
  andrelated algorithms},\ }\href@noop {} {\bibfield  {journal} {\bibinfo
  {journal} {Advanced Quantum Technologies}\ } (\bibinfo {year}
  {2020})}\BibitemShut {NoStop}%
\bibitem [{\citenamefont {Ding}\ \emph {et~al.}(2022)\citenamefont {Ding},
  \citenamefont {Xu}, \citenamefont {Zhang}, \citenamefont {Huang},\ and\
  \citenamefont {Bao}}]{ExpMMD2022}%
  \BibitemOpen
  \bibfield  {author} {\bibinfo {author} {\bibfnamefont {C.}~\bibnamefont
  {Ding}}, \bibinfo {author} {\bibfnamefont {X.-Y.}\ \bibnamefont {Xu}},
  \bibinfo {author} {\bibfnamefont {S.}~\bibnamefont {Zhang}}, \bibinfo
  {author} {\bibfnamefont {H.-L.}\ \bibnamefont {Huang}},\ and\ \bibinfo
  {author} {\bibfnamefont {W.-S.}\ \bibnamefont {Bao}},\ }\bibfield  {title}
  {\bibinfo {title} {Evaluating the resilience of variational quantum
  algorithms to leakage noise},\ }\href@noop {} {\bibfield  {journal} {\bibinfo
   {journal} {Physical Review A}\ }\textbf {\bibinfo {volume} {106}},\ \bibinfo
  {pages} {042421} (\bibinfo {year} {2022})}\BibitemShut {NoStop}%
\bibitem [{\citenamefont {Nakaji}\ and\ \citenamefont
  {Yamamoto}(2021)}]{AlterExpressibility2021}%
  \BibitemOpen
  \bibfield  {author} {\bibinfo {author} {\bibfnamefont {K.}~\bibnamefont
  {Nakaji}}\ and\ \bibinfo {author} {\bibfnamefont {N.}~\bibnamefont
  {Yamamoto}},\ }\bibfield  {title} {\bibinfo {title} {Expressibility of the
  alternating layered ansatz for quantum computation},\ }\href@noop {}
  {\bibfield  {journal} {\bibinfo  {journal} {Quantum}\ }\textbf {\bibinfo
  {volume} {5}},\ \bibinfo {pages} {434} (\bibinfo {year} {2021})}\BibitemShut
  {NoStop}%
\bibitem [{\citenamefont {Mao}\ \emph {et~al.}(2023)\citenamefont {Mao},
  \citenamefont {Shresthamali},\ and\ \citenamefont {Kondo}}]{Mao2023zkt}%
  \BibitemOpen
  \bibfield  {author} {\bibinfo {author} {\bibfnamefont {Y.}~\bibnamefont
  {Mao}}, \bibinfo {author} {\bibfnamefont {S.}~\bibnamefont {Shresthamali}},\
  and\ \bibinfo {author} {\bibfnamefont {M.}~\bibnamefont {Kondo}},\ }\bibfield
   {title} {\bibinfo {title} {Quantum circuit fidelity improvement with long
  short-term memory networks},\ }\href@noop {} {\  (\bibinfo {year}
  {2023})}\BibitemShut {NoStop}%
\bibitem [{\citenamefont {Altares-L{\'o}pez}\ \emph {et~al.}(2024)\citenamefont
  {Altares-L{\'o}pez}, \citenamefont {Garc{\'\i}a-Ripoll},\ and\ \citenamefont
  {Ribeiro}}]{altares2024autoqml}%
  \BibitemOpen
  \bibfield  {author} {\bibinfo {author} {\bibfnamefont {S.}~\bibnamefont
  {Altares-L{\'o}pez}}, \bibinfo {author} {\bibfnamefont {J.~J.}\ \bibnamefont
  {Garc{\'\i}a-Ripoll}},\ and\ \bibinfo {author} {\bibfnamefont
  {A.}~\bibnamefont {Ribeiro}},\ }\bibfield  {title} {\bibinfo {title}
  {Auto{QML}: Automatic generation and training of robust quantum-inspired
  classifiers by using evolutionary algorithms on grayscale images},\
  }\href@noop {} {\bibfield  {journal} {\bibinfo  {journal} {Expert Systems
  with Applications}\ }\textbf {\bibinfo {volume} {244}},\ \bibinfo {pages}
  {122984} (\bibinfo {year} {2024})}\BibitemShut {NoStop}%
\bibitem [{\citenamefont {He}\ \emph {et~al.}(2023)\citenamefont {He},
  \citenamefont {Deng}, \citenamefont {Zheng}, \citenamefont {Li},\ and\
  \citenamefont {Situ}}]{GSQAS2023}%
  \BibitemOpen
  \bibfield  {author} {\bibinfo {author} {\bibfnamefont {Z.}~\bibnamefont
  {He}}, \bibinfo {author} {\bibfnamefont {M.}~\bibnamefont {Deng}}, \bibinfo
  {author} {\bibfnamefont {S.}~\bibnamefont {Zheng}}, \bibinfo {author}
  {\bibfnamefont {L.}~\bibnamefont {Li}},\ and\ \bibinfo {author}
  {\bibfnamefont {H.}~\bibnamefont {Situ}},\ }\bibfield  {title} {\bibinfo
  {title} {{GSQAS}: {G}raph {S}elf-supervised {Q}uantum {A}rchitecture
  {S}earch},\ }\href@noop {} {\bibfield  {journal} {\bibinfo  {journal}
  {Physica A: Statistical Mechanics and its Applications}\ }\textbf {\bibinfo
  {volume} {630}},\ \bibinfo {pages} {129286} (\bibinfo {year}
  {2023})}\BibitemShut {NoStop}%
\bibitem [{\citenamefont {Xiao}\ and\ \citenamefont
  {Zhu}(2023)}]{introductiontransformers2023}%
  \BibitemOpen
  \bibfield  {author} {\bibinfo {author} {\bibfnamefont {T.}~\bibnamefont
  {Xiao}}\ and\ \bibinfo {author} {\bibfnamefont {J.}~\bibnamefont {Zhu}},\
  }\href@noop {} {\bibinfo {title} {Introduction to transformers: an nlp
  perspective}} (\bibinfo {year} {2023}),\ \Eprint
  {https://arxiv.org/abs/2311.17633} {2311.17633} \BibitemShut {NoStop}%
\bibitem [{\citenamefont {Apak}\ \emph {et~al.}(2024)\citenamefont {Apak},
  \citenamefont {Bandic}, \citenamefont {Sarkar},\ and\ \citenamefont
  {Feld}}]{apak2024ketgpt}%
  \BibitemOpen
  \bibfield  {author} {\bibinfo {author} {\bibfnamefont {B.}~\bibnamefont
  {Apak}}, \bibinfo {author} {\bibfnamefont {M.}~\bibnamefont {Bandic}},
  \bibinfo {author} {\bibfnamefont {A.}~\bibnamefont {Sarkar}},\ and\ \bibinfo
  {author} {\bibfnamefont {S.}~\bibnamefont {Feld}},\ }\bibfield  {title}
  {\bibinfo {title} {Ketgpt--dataset augmentation of quantum circuits using
  transformers},\ }in\ \href@noop {} {\emph {\bibinfo {booktitle}
  {International Conference on Computational Science}}}\ (\bibinfo
  {organization} {Springer},\ \bibinfo {year} {2024})\ pp.\ \bibinfo {pages}
  {235--251}\BibitemShut {NoStop}%
\bibitem [{\citenamefont {Zhang}\ and\ \citenamefont
  {Ventra}(2023)}]{Zhang2022TransformerQS}%
  \BibitemOpen
  \bibfield  {author} {\bibinfo {author} {\bibfnamefont {Y.}~\bibnamefont
  {Zhang}}\ and\ \bibinfo {author} {\bibfnamefont {M.~D.}\ \bibnamefont
  {Ventra}},\ }\bibfield  {title} {\bibinfo {title} {Transformer quantum state:
  A multipurpose model for quantum many-body problems},\ }\href@noop {}
  {\bibfield  {journal} {\bibinfo  {journal} {Physical Review B}\ ,\ \bibinfo
  {pages} {075147}} (\bibinfo {year} {2023})}\BibitemShut {NoStop}%
\bibitem [{\citenamefont {Wang}\ \emph {et~al.}(2022)\citenamefont {Wang},
  \citenamefont {Liu}, \citenamefont {Cheng}, \citenamefont {Liang},
  \citenamefont {Gu}, \citenamefont {Li}, \citenamefont {Ding}, \citenamefont
  {Jiang}, \citenamefont {Shi}, \citenamefont {Qian}, \citenamefont {Pan},
  \citenamefont {Chong},\ and\ \citenamefont {Han}}]{QuEst2022}%
  \BibitemOpen
  \bibfield  {author} {\bibinfo {author} {\bibfnamefont {H.}~\bibnamefont
  {Wang}}, \bibinfo {author} {\bibfnamefont {P.}~\bibnamefont {Liu}}, \bibinfo
  {author} {\bibfnamefont {J.}~\bibnamefont {Cheng}}, \bibinfo {author}
  {\bibfnamefont {Z.}~\bibnamefont {Liang}}, \bibinfo {author} {\bibfnamefont
  {J.}~\bibnamefont {Gu}}, \bibinfo {author} {\bibfnamefont {Z.}~\bibnamefont
  {Li}}, \bibinfo {author} {\bibfnamefont {Y.}~\bibnamefont {Ding}}, \bibinfo
  {author} {\bibfnamefont {W.}~\bibnamefont {Jiang}}, \bibinfo {author}
  {\bibfnamefont {Y.}~\bibnamefont {Shi}}, \bibinfo {author} {\bibfnamefont
  {X.}~\bibnamefont {Qian}}, \bibinfo {author} {\bibfnamefont {D.}~\bibnamefont
  {Pan}}, \bibinfo {author} {\bibfnamefont {F.}~\bibnamefont {Chong}},\ and\
  \bibinfo {author} {\bibfnamefont {S.}~\bibnamefont {Han}},\ }\bibfield
  {title} {\bibinfo {title} {Quest: Graph transformer for quantum circuit
  reliability estimation},\ }in\ \href@noop {} {\emph {\bibinfo {booktitle}
  {IEEE/ACM International Conference on Computer-Aided Design (ICCAD)}}}\
  (\bibinfo {year} {2022})\BibitemShut {NoStop}%
\bibitem [{\citenamefont {Nguyen}\ \emph {et~al.}(2024)\citenamefont {Nguyen},
  \citenamefont {Nguyen}, \citenamefont {Chen}, \citenamefont {Khan},
  \citenamefont {Churchill},\ and\ \citenamefont
  {Luu}}]{nguyen2024qclusformer}%
  \BibitemOpen
  \bibfield  {author} {\bibinfo {author} {\bibfnamefont {X.-B.}\ \bibnamefont
  {Nguyen}}, \bibinfo {author} {\bibfnamefont {H.-Q.}\ \bibnamefont {Nguyen}},
  \bibinfo {author} {\bibfnamefont {S.~Y.-C.}\ \bibnamefont {Chen}}, \bibinfo
  {author} {\bibfnamefont {S.~U.}\ \bibnamefont {Khan}}, \bibinfo {author}
  {\bibfnamefont {H.}~\bibnamefont {Churchill}},\ and\ \bibinfo {author}
  {\bibfnamefont {K.}~\bibnamefont {Luu}},\ }\bibfield  {title} {\bibinfo
  {title} {Qclusformer: A quantum transformer-based framework for unsupervised
  visual clustering},\ }\href@noop {} {\bibfield  {journal} {\bibinfo
  {journal} {arXiv preprint arXiv:2405.19722}\ } (\bibinfo {year}
  {2024})}\BibitemShut {NoStop}%
\end{thebibliography}%


\begin{thebibliography}{24}%
\makeatletter
\providecommand \@ifxundefined [1]{%
 \@ifx{#1\undefined}
}%
\providecommand \@ifnum [1]{%
 \ifnum #1\expandafter \@firstoftwo
 \else \expandafter \@secondoftwo
 \fi
}%
\providecommand \@ifx [1]{%
 \ifx #1\expandafter \@firstoftwo
 \else \expandafter \@secondoftwo
 \fi
}%
\providecommand \natexlab [1]{#1}%
\providecommand \enquote  [1]{``#1''}%
\providecommand \bibnamefont  [1]{#1}%
\providecommand \bibfnamefont [1]{#1}%
\providecommand \citenamefont [1]{#1}%
\providecommand \href@noop [0]{\@secondoftwo}%
\providecommand \href [0]{\begingroup \@sanitize@url \@href}%
\providecommand \@href[1]{\@@startlink{#1}\@@href}%
\providecommand \@@href[1]{\endgroup#1\@@endlink}%
\providecommand \@sanitize@url [0]{\catcode `\\12\catcode `\$12\catcode
  `\&12\catcode `\#12\catcode `\^12\catcode `\_12\catcode `\%12\relax}%
\providecommand \@@startlink[1]{}%
\providecommand \@@endlink[0]{}%
\providecommand \url  [0]{\begingroup\@sanitize@url \@url }%
\providecommand \@url [1]{\endgroup\@href {#1}{\urlprefix }}%
\providecommand \urlprefix  [0]{URL }%
\providecommand \Eprint [0]{\href }%
\providecommand \doibase [0]{https://doi.org/}%
\providecommand \selectlanguage [0]{\@gobble}%
\providecommand \bibinfo  [0]{\@secondoftwo}%
\providecommand \bibfield  [0]{\@secondoftwo}%
\providecommand \translation [1]{[#1]}%
\providecommand \BibitemOpen [0]{}%
\providecommand \bibitemStop [0]{}%
\providecommand \bibitemNoStop [0]{.\EOS\space}%
\providecommand \EOS [0]{\spacefactor3000\relax}%
\providecommand \BibitemShut  [1]{\csname bibitem#1\endcsname}%
\let\auto@bib@innerbib\@empty
\bibitem [{\citenamefont {Endo}\ \emph {et~al.}(2020)\citenamefont {Endo},
  \citenamefont {Sun}, \citenamefont {Li}, \citenamefont {Benjamin},\ and\
  \citenamefont {Yuan}}]{endo2020variational}%
  \BibitemOpen
  \bibfield  {author} {\bibinfo {author} {\bibfnamefont {S.}~\bibnamefont
  {Endo}}, \bibinfo {author} {\bibfnamefont {J.}~\bibnamefont {Sun}}, \bibinfo
  {author} {\bibfnamefont {Y.}~\bibnamefont {Li}}, \bibinfo {author}
  {\bibfnamefont {S.~C.}\ \bibnamefont {Benjamin}},\ and\ \bibinfo {author}
  {\bibfnamefont {X.}~\bibnamefont {Yuan}},\ }\bibfield  {title} {\bibinfo
  {title} {Variational quantum simulation of general processes},\ }\href@noop
  {} {\bibfield  {journal} {\bibinfo  {journal} {Physical Review Letters}\
  }\textbf {\bibinfo {volume} {125}},\ \bibinfo {pages} {010501} (\bibinfo
  {year} {2020})}\BibitemShut {NoStop}%
\bibitem [{\citenamefont {Mauranyapin}\ \emph {et~al.}(2022)\citenamefont
  {Mauranyapin}, \citenamefont {Terrasson},\ and\ \citenamefont
  {Bowen}}]{QuBiotechnology2022}%
  \BibitemOpen
  \bibfield  {author} {\bibinfo {author} {\bibfnamefont {N.~P.}\ \bibnamefont
  {Mauranyapin}}, \bibinfo {author} {\bibfnamefont {A.}~\bibnamefont
  {Terrasson}},\ and\ \bibinfo {author} {\bibfnamefont {W.~P.}\ \bibnamefont
  {Bowen}},\ }\bibfield  {title} {\bibinfo {title} {Quantum biotechnology},\
  }\href@noop {} {\bibfield  {journal} {\bibinfo  {journal} {Advanced Quantum
  Technologies}\ }\textbf {\bibinfo {volume} {5}},\ \bibinfo {pages} {2100139}
  (\bibinfo {year} {2022})}\BibitemShut {NoStop}%
\bibitem [{\citenamefont {Cerezo}\ \emph {et~al.}(2021)\citenamefont {Cerezo},
  \citenamefont {Arrasmith}, \citenamefont {Babbush}, \citenamefont {Benjamin},
  \citenamefont {Endo}, \citenamefont {Fujii}, \citenamefont {McClean},
  \citenamefont {Mitarai}, \citenamefont {Yuan}, \citenamefont {Cincio} \emph
  {et~al.}}]{cerezo2021variational}%
  \BibitemOpen
  \bibfield  {author} {\bibinfo {author} {\bibfnamefont {M.}~\bibnamefont
  {Cerezo}}, \bibinfo {author} {\bibfnamefont {A.}~\bibnamefont {Arrasmith}},
  \bibinfo {author} {\bibfnamefont {R.}~\bibnamefont {Babbush}}, \bibinfo
  {author} {\bibfnamefont {S.~C.}\ \bibnamefont {Benjamin}}, \bibinfo {author}
  {\bibfnamefont {S.}~\bibnamefont {Endo}}, \bibinfo {author} {\bibfnamefont
  {K.}~\bibnamefont {Fujii}}, \bibinfo {author} {\bibfnamefont {J.~R.}\
  \bibnamefont {McClean}}, \bibinfo {author} {\bibfnamefont {K.}~\bibnamefont
  {Mitarai}}, \bibinfo {author} {\bibfnamefont {X.}~\bibnamefont {Yuan}},
  \bibinfo {author} {\bibfnamefont {L.}~\bibnamefont {Cincio}}, \emph
  {et~al.},\ }\bibfield  {title} {\bibinfo {title} {Variational quantum
  algorithms},\ }\href@noop {} {\bibfield  {journal} {\bibinfo  {journal}
  {Nature Reviews Physics}\ }\textbf {\bibinfo {volume} {3}},\ \bibinfo {pages}
  {625} (\bibinfo {year} {2021})}\BibitemShut {NoStop}%
\bibitem [{\citenamefont {Situ}\ \emph {et~al.}(2020)\citenamefont {Situ},
  \citenamefont {He}, \citenamefont {Wang}, \citenamefont {Li},\ and\
  \citenamefont {Zheng}}]{situ2020quantum}%
  \BibitemOpen
  \bibfield  {author} {\bibinfo {author} {\bibfnamefont {H.}~\bibnamefont
  {Situ}}, \bibinfo {author} {\bibfnamefont {Z.}~\bibnamefont {He}}, \bibinfo
  {author} {\bibfnamefont {Y.}~\bibnamefont {Wang}}, \bibinfo {author}
  {\bibfnamefont {L.}~\bibnamefont {Li}},\ and\ \bibinfo {author}
  {\bibfnamefont {S.}~\bibnamefont {Zheng}},\ }\bibfield  {title} {\bibinfo
  {title} {Quantum generative adversarial network for generating discrete
  distribution},\ }\href@noop {} {\bibfield  {journal} {\bibinfo  {journal}
  {Information Sciences}\ }\textbf {\bibinfo {volume} {538}},\ \bibinfo {pages}
  {193} (\bibinfo {year} {2020})}\BibitemShut {NoStop}%
\bibitem [{\citenamefont {Pan}\ \emph {et~al.}(2023)\citenamefont {Pan},
  \citenamefont {Lu}, \citenamefont {Wang}, \citenamefont {Hua}, \citenamefont
  {Xu}, \citenamefont {Li}, \citenamefont {Cai}, \citenamefont {Li},
  \citenamefont {Wang}, \citenamefont {Song}, \citenamefont {Zou},
  \citenamefont {Deng},\ and\ \citenamefont {Sun}}]{pan2023deep}%
  \BibitemOpen
  \bibfield  {author} {\bibinfo {author} {\bibfnamefont {X.}~\bibnamefont
  {Pan}}, \bibinfo {author} {\bibfnamefont {Z.}~\bibnamefont {Lu}}, \bibinfo
  {author} {\bibfnamefont {W.}~\bibnamefont {Wang}}, \bibinfo {author}
  {\bibfnamefont {Z.}~\bibnamefont {Hua}}, \bibinfo {author} {\bibfnamefont
  {Y.}~\bibnamefont {Xu}}, \bibinfo {author} {\bibfnamefont {W.}~\bibnamefont
  {Li}}, \bibinfo {author} {\bibfnamefont {W.}~\bibnamefont {Cai}}, \bibinfo
  {author} {\bibfnamefont {X.}~\bibnamefont {Li}}, \bibinfo {author}
  {\bibfnamefont {H.}~\bibnamefont {Wang}}, \bibinfo {author} {\bibfnamefont
  {Y.-P.}\ \bibnamefont {Song}}, \bibinfo {author} {\bibfnamefont {C.-L.}\
  \bibnamefont {Zou}}, \bibinfo {author} {\bibfnamefont {D.-L.}\ \bibnamefont
  {Deng}},\ and\ \bibinfo {author} {\bibfnamefont {L.}~\bibnamefont {Sun}},\
  }\bibfield  {title} {\bibinfo {title} {Deep quantum neural networks on a
  superconducting processor},\ }\href@noop {} {\bibfield  {journal} {\bibinfo
  {journal} {Nature Communications}\ }\textbf {\bibinfo {volume} {14}},\
  \bibinfo {pages} {4006} (\bibinfo {year} {2023})}\BibitemShut {NoStop}%
\bibitem [{\citenamefont {Ding}\ \emph {et~al.}(2023)\citenamefont {Ding},
  \citenamefont {Xu}, \citenamefont {Niu}, \citenamefont {Zhang}, \citenamefont
  {Huang},\ and\ \citenamefont {Bao}}]{ding2023active}%
  \BibitemOpen
  \bibfield  {author} {\bibinfo {author} {\bibfnamefont {C.}~\bibnamefont
  {Ding}}, \bibinfo {author} {\bibfnamefont {X.-Y.}\ \bibnamefont {Xu}},
  \bibinfo {author} {\bibfnamefont {Y.-F.}\ \bibnamefont {Niu}}, \bibinfo
  {author} {\bibfnamefont {S.}~\bibnamefont {Zhang}}, \bibinfo {author}
  {\bibfnamefont {H.-L.}\ \bibnamefont {Huang}},\ and\ \bibinfo {author}
  {\bibfnamefont {W.-S.}\ \bibnamefont {Bao}},\ }\bibfield  {title} {\bibinfo
  {title} {Active learning on a programmable photonic quantum processor},\
  }\href@noop {} {\bibfield  {journal} {\bibinfo  {journal} {Quantum Science
  and Technology}\ }\textbf {\bibinfo {volume} {8}},\ \bibinfo {pages} {035030}
  (\bibinfo {year} {2023})}\BibitemShut {NoStop}%
\bibitem [{\citenamefont {Shi}\ \emph {et~al.}(2023)\citenamefont {Shi},
  \citenamefont {Wang}, \citenamefont {Lou}, \citenamefont {Zhang},\ and\
  \citenamefont {Li}}]{shi2023parameterized}%
  \BibitemOpen
  \bibfield  {author} {\bibinfo {author} {\bibfnamefont {J.}~\bibnamefont
  {Shi}}, \bibinfo {author} {\bibfnamefont {W.}~\bibnamefont {Wang}}, \bibinfo
  {author} {\bibfnamefont {X.}~\bibnamefont {Lou}}, \bibinfo {author}
  {\bibfnamefont {S.}~\bibnamefont {Zhang}},\ and\ \bibinfo {author}
  {\bibfnamefont {X.}~\bibnamefont {Li}},\ }\bibfield  {title} {\bibinfo
  {title} {Parameterized {H}amiltonian learning with quantum circuit},\
  }\href@noop {} {\bibfield  {journal} {\bibinfo  {journal} {IEEE Transactions
  on Pattern Analysis and Machine Intelligence}\ }\textbf {\bibinfo {volume}
  {45}},\ \bibinfo {pages} {6086} (\bibinfo {year} {2023})}\BibitemShut
  {NoStop}%
\bibitem [{\citenamefont {Ni}\ \emph {et~al.}(2024)\citenamefont {Ni},
  \citenamefont {Cai}, \citenamefont {Liu}, \citenamefont {Qin}, \citenamefont
  {Gao},\ and\ \citenamefont {Wen}}]{ni2024multilevel}%
  \BibitemOpen
  \bibfield  {author} {\bibinfo {author} {\bibfnamefont {X.-H.}\ \bibnamefont
  {Ni}}, \bibinfo {author} {\bibfnamefont {B.-B.}\ \bibnamefont {Cai}},
  \bibinfo {author} {\bibfnamefont {H.-L.}\ \bibnamefont {Liu}}, \bibinfo
  {author} {\bibfnamefont {S.-J.}\ \bibnamefont {Qin}}, \bibinfo {author}
  {\bibfnamefont {F.}~\bibnamefont {Gao}},\ and\ \bibinfo {author}
  {\bibfnamefont {Q.-Y.}\ \bibnamefont {Wen}},\ }\bibfield  {title} {\bibinfo
  {title} {Multilevel leapfrogging initialization strategy for quantum
  approximate optimization algorithm},\ }\href@noop {} {\bibfield  {journal}
  {\bibinfo  {journal} {Advanced Quantum Technologies}\ }\textbf {\bibinfo
  {volume} {7}},\ \bibinfo {pages} {2300419} (\bibinfo {year}
  {2024})}\BibitemShut {NoStop}%
\bibitem [{\citenamefont {Martyniuk}\ \emph {et~al.}(2024)\citenamefont
  {Martyniuk}, \citenamefont {Jung},\ and\ \citenamefont
  {Paschke}}]{martyniuk2024quantum}%
  \BibitemOpen
  \bibfield  {author} {\bibinfo {author} {\bibfnamefont {D.}~\bibnamefont
  {Martyniuk}}, \bibinfo {author} {\bibfnamefont {J.}~\bibnamefont {Jung}},\
  and\ \bibinfo {author} {\bibfnamefont {A.}~\bibnamefont {Paschke}},\
  }\bibfield  {title} {\bibinfo {title} {Quantum architecture search: a
  survey},\ }\href@noop {} {\bibfield  {journal} {\bibinfo  {journal} {arXiv
  preprint arXiv:2406.06210}\ } (\bibinfo {year} {2024})}\BibitemShut {NoStop}%
\bibitem [{\citenamefont {Anagolum}\ \emph {et~al.}(2024)\citenamefont
  {Anagolum}, \citenamefont {Alavisamani}, \citenamefont {Das}, \citenamefont
  {Qureshi},\ and\ \citenamefont {Shi}}]{anagolum2024elivagar}%
  \BibitemOpen
  \bibfield  {author} {\bibinfo {author} {\bibfnamefont {S.}~\bibnamefont
  {Anagolum}}, \bibinfo {author} {\bibfnamefont {N.}~\bibnamefont
  {Alavisamani}}, \bibinfo {author} {\bibfnamefont {P.}~\bibnamefont {Das}},
  \bibinfo {author} {\bibfnamefont {M.}~\bibnamefont {Qureshi}},\ and\ \bibinfo
  {author} {\bibfnamefont {Y.}~\bibnamefont {Shi}},\ }\bibfield  {title}
  {\bibinfo {title} {{\'E}liv{\'a}gar: Efficient quantum circuit search for
  classification},\ }in\ \href@noop {} {\emph {\bibinfo {booktitle}
  {Proceedings of the 29th ACM International Conference on Architectural
  Support for Programming Languages and Operating Systems, Volume 2}}}\
  (\bibinfo {year} {2024})\ pp.\ \bibinfo {pages} {336--353}\BibitemShut
  {NoStop}%
\bibitem [{\citenamefont {He}\ \emph {et~al.}(2024)\citenamefont {He},
  \citenamefont {Deng}, \citenamefont {Zheng}, \citenamefont {Li},\ and\
  \citenamefont {Situ}}]{zhimin2024aaai}%
  \BibitemOpen
  \bibfield  {author} {\bibinfo {author} {\bibfnamefont {Z.}~\bibnamefont
  {He}}, \bibinfo {author} {\bibfnamefont {M.}~\bibnamefont {Deng}}, \bibinfo
  {author} {\bibfnamefont {S.}~\bibnamefont {Zheng}}, \bibinfo {author}
  {\bibfnamefont {L.}~\bibnamefont {Li}},\ and\ \bibinfo {author}
  {\bibfnamefont {H.}~\bibnamefont {Situ}},\ }\bibfield  {title} {\bibinfo
  {title} {Training-free quantum architecture search},\ }in\ \href@noop {}
  {\emph {\bibinfo {booktitle} {Proceedings of the AAAI Conference on
  Artificial Intelligence}}},\ Vol.~\bibinfo {volume} {38}\ (\bibinfo {year}
  {2024})\ pp.\ \bibinfo {pages} {12430--12438}\BibitemShut {NoStop}%
\bibitem [{\citenamefont {Situ}\ \emph {et~al.}(2024)\citenamefont {Situ},
  \citenamefont {He}, \citenamefont {Zheng},\ and\ \citenamefont
  {Li}}]{situ2024distributed}%
  \BibitemOpen
  \bibfield  {author} {\bibinfo {author} {\bibfnamefont {H.}~\bibnamefont
  {Situ}}, \bibinfo {author} {\bibfnamefont {Z.}~\bibnamefont {He}}, \bibinfo
  {author} {\bibfnamefont {S.}~\bibnamefont {Zheng}},\ and\ \bibinfo {author}
  {\bibfnamefont {L.}~\bibnamefont {Li}},\ }\bibfield  {title} {\bibinfo
  {title} {Distributed quantum architecture search},\ }\href@noop {} {\bibfield
   {journal} {\bibinfo  {journal} {Physical Review A}\ }\textbf {\bibinfo
  {volume} {110}},\ \bibinfo {pages} {022403} (\bibinfo {year}
  {2024})}\BibitemShut {NoStop}%
\bibitem [{\citenamefont {Sim}\ \emph {et~al.}(2019)\citenamefont {Sim},
  \citenamefont {Johnson},\ and\ \citenamefont
  {Aspuru-Guzik}}]{HybridExpressibility2019}%
  \BibitemOpen
  \bibfield  {author} {\bibinfo {author} {\bibfnamefont {S.}~\bibnamefont
  {Sim}}, \bibinfo {author} {\bibfnamefont {P.~D.}\ \bibnamefont {Johnson}},\
  and\ \bibinfo {author} {\bibfnamefont {A.}~\bibnamefont {Aspuru-Guzik}},\
  }\bibfield  {title} {\bibinfo {title} {Expressibility and entangling
  capability of parameterized quantum circuits for hybrid quantum-classical
  algorithms},\ }\href@noop {} {\bibfield  {journal} {\bibinfo  {journal}
  {Advanced Quantum Technologies}\ }\textbf {\bibinfo {volume} {2}},\ \bibinfo
  {pages} {1900070} (\bibinfo {year} {2019})}\BibitemShut {NoStop}%
\bibitem [{\citenamefont {Rasmussen}\ \emph {et~al.}(2020)\citenamefont
  {Rasmussen}, \citenamefont {Loft}, \citenamefont {B{\ae}kkegaard},
  \citenamefont {Kues},\ and\ \citenamefont {Zinner}}]{KL_R2020}%
  \BibitemOpen
  \bibfield  {author} {\bibinfo {author} {\bibfnamefont {S.~E.}\ \bibnamefont
  {Rasmussen}}, \bibinfo {author} {\bibfnamefont {N.~J.~S.}\ \bibnamefont
  {Loft}}, \bibinfo {author} {\bibfnamefont {T.}~\bibnamefont
  {B{\ae}kkegaard}}, \bibinfo {author} {\bibfnamefont {M.}~\bibnamefont
  {Kues}},\ and\ \bibinfo {author} {\bibfnamefont {N.~T.}\ \bibnamefont
  {Zinner}},\ }\bibfield  {title} {\bibinfo {title} {Reducing the amount of
  single-qubit rotations in vqe and related algorithms},\ }\href@noop {}
  {\bibfield  {journal} {\bibinfo  {journal} {Advanced Quantum Technologies}\
  }\textbf {\bibinfo {volume} {3}},\ \bibinfo {pages} {2000063} (\bibinfo
  {year} {2020})}\BibitemShut {NoStop}%
\bibitem [{\citenamefont {Ding}\ \emph {et~al.}(2022)\citenamefont {Ding},
  \citenamefont {Xu}, \citenamefont {Zhang}, \citenamefont {Huang},\ and\
  \citenamefont {Bao}}]{ExpMMD2022}%
  \BibitemOpen
  \bibfield  {author} {\bibinfo {author} {\bibfnamefont {C.}~\bibnamefont
  {Ding}}, \bibinfo {author} {\bibfnamefont {X.-Y.}\ \bibnamefont {Xu}},
  \bibinfo {author} {\bibfnamefont {S.}~\bibnamefont {Zhang}}, \bibinfo
  {author} {\bibfnamefont {H.-L.}\ \bibnamefont {Huang}},\ and\ \bibinfo
  {author} {\bibfnamefont {W.-S.}\ \bibnamefont {Bao}},\ }\bibfield  {title}
  {\bibinfo {title} {Evaluating the resilience of variational quantum
  algorithms to leakage noise},\ }\href@noop {} {\bibfield  {journal} {\bibinfo
   {journal} {Physical Review A}\ }\textbf {\bibinfo {volume} {106}},\ \bibinfo
  {pages} {042421} (\bibinfo {year} {2022})}\BibitemShut {NoStop}%
\bibitem [{\citenamefont {Zhang}\ \emph {et~al.}(2021)\citenamefont {Zhang},
  \citenamefont {Hsieh}, \citenamefont {Zhang},\ and\ \citenamefont
  {Yao}}]{zhangHsieh2021}%
  \BibitemOpen
  \bibfield  {author} {\bibinfo {author} {\bibfnamefont {S.-X.}\ \bibnamefont
  {Zhang}}, \bibinfo {author} {\bibfnamefont {C.-Y.}\ \bibnamefont {Hsieh}},
  \bibinfo {author} {\bibfnamefont {S.}~\bibnamefont {Zhang}},\ and\ \bibinfo
  {author} {\bibfnamefont {H.}~\bibnamefont {Yao}},\ }\bibfield  {title}
  {\bibinfo {title} {Neural predictor based quantum architecture search},\
  }\href@noop {} {\bibfield  {journal} {\bibinfo  {journal} {Machine Learning:
  Science and Technology}\ }\textbf {\bibinfo {volume} {2}},\ \bibinfo {pages}
  {045027} (\bibinfo {year} {2021})}\BibitemShut {NoStop}%
\bibitem [{\citenamefont {Mao}\ \emph {et~al.}(2023)\citenamefont {Mao},
  \citenamefont {Shresthamali},\ and\ \citenamefont {Kondo}}]{Mao2023zkt}%
  \BibitemOpen
  \bibfield  {author} {\bibinfo {author} {\bibfnamefont {Y.}~\bibnamefont
  {Mao}}, \bibinfo {author} {\bibfnamefont {S.}~\bibnamefont {Shresthamali}},\
  and\ \bibinfo {author} {\bibfnamefont {M.}~\bibnamefont {Kondo}},\ }\bibfield
   {title} {\bibinfo {title} {Quantum circuit fidelity improvement with long
  short-term memory networks},\ }\href@noop {} {\bibfield  {journal} {\bibinfo
  {journal} {arXiv preprint arXiv:2303.17523}\ } (\bibinfo {year}
  {2023})}\BibitemShut {NoStop}%
\bibitem [{\citenamefont {Altares-L{\'o}pez}\ \emph {et~al.}(2021)\citenamefont
  {Altares-L{\'o}pez}, \citenamefont {Ribeiro},\ and\ \citenamefont
  {Garc{\'\i}a-Ripoll}}]{altares2021automatic}%
  \BibitemOpen
  \bibfield  {author} {\bibinfo {author} {\bibfnamefont {S.}~\bibnamefont
  {Altares-L{\'o}pez}}, \bibinfo {author} {\bibfnamefont {A.}~\bibnamefont
  {Ribeiro}},\ and\ \bibinfo {author} {\bibfnamefont {J.~J.}\ \bibnamefont
  {Garc{\'\i}a-Ripoll}},\ }\bibfield  {title} {\bibinfo {title} {Automatic
  design of quantum feature maps},\ }\href@noop {} {\bibfield  {journal}
  {\bibinfo  {journal} {Quantum Science and Technology}\ }\textbf {\bibinfo
  {volume} {6}},\ \bibinfo {pages} {045015} (\bibinfo {year}
  {2021})}\BibitemShut {NoStop}%
\bibitem [{\citenamefont {He}\ \emph {et~al.}(2023)\citenamefont {He},
  \citenamefont {Deng}, \citenamefont {Zheng}, \citenamefont {Li},\ and\
  \citenamefont {Situ}}]{GSQAS2023}%
  \BibitemOpen
  \bibfield  {author} {\bibinfo {author} {\bibfnamefont {Z.}~\bibnamefont
  {He}}, \bibinfo {author} {\bibfnamefont {M.}~\bibnamefont {Deng}}, \bibinfo
  {author} {\bibfnamefont {S.}~\bibnamefont {Zheng}}, \bibinfo {author}
  {\bibfnamefont {L.}~\bibnamefont {Li}},\ and\ \bibinfo {author}
  {\bibfnamefont {H.}~\bibnamefont {Situ}},\ }\bibfield  {title} {\bibinfo
  {title} {{GSQAS}: {G}raph self-supervised quantum architecture search},\
  }\href@noop {} {\bibfield  {journal} {\bibinfo  {journal} {Physica A:
  Statistical Mechanics and its Applications}\ }\textbf {\bibinfo {volume}
  {630}},\ \bibinfo {pages} {129286} (\bibinfo {year} {2023})}\BibitemShut
  {NoStop}%
\bibitem [{\citenamefont {Xiao}\ and\ \citenamefont
  {Zhu}(2023)}]{introductiontransformers2023}%
  \BibitemOpen
  \bibfield  {author} {\bibinfo {author} {\bibfnamefont {T.}~\bibnamefont
  {Xiao}}\ and\ \bibinfo {author} {\bibfnamefont {J.}~\bibnamefont {Zhu}},\
  }\bibfield  {title} {\bibinfo {title} {Introduction to transformers: an {NLP}
  perspective},\ }\href@noop {} {\bibfield  {journal} {\bibinfo  {journal}
  {arXiv preprint arXiv:2311.17633}\ } (\bibinfo {year} {2023})}\BibitemShut
  {NoStop}%
\bibitem [{\citenamefont {Apak}\ \emph {et~al.}(2024)\citenamefont {Apak},
  \citenamefont {Bandic}, \citenamefont {Sarkar},\ and\ \citenamefont
  {Feld}}]{apak2024ketgpt}%
  \BibitemOpen
  \bibfield  {author} {\bibinfo {author} {\bibfnamefont {B.}~\bibnamefont
  {Apak}}, \bibinfo {author} {\bibfnamefont {M.}~\bibnamefont {Bandic}},
  \bibinfo {author} {\bibfnamefont {A.}~\bibnamefont {Sarkar}},\ and\ \bibinfo
  {author} {\bibfnamefont {S.}~\bibnamefont {Feld}},\ }\bibfield  {title}
  {\bibinfo {title} {Ketgpt--dataset augmentation of quantum circuits using
  transformers},\ }in\ \href@noop {} {\emph {\bibinfo {booktitle}
  {International Conference on Computational Science}}}\ (\bibinfo
  {organization} {Springer},\ \bibinfo {year} {2024})\ pp.\ \bibinfo {pages}
  {235--251}\BibitemShut {NoStop}%
\bibitem [{\citenamefont {Zhang}\ and\ \citenamefont
  {Ventra}(2023)}]{Zhang2022TransformerQS}%
  \BibitemOpen
  \bibfield  {author} {\bibinfo {author} {\bibfnamefont {Y.}~\bibnamefont
  {Zhang}}\ and\ \bibinfo {author} {\bibfnamefont {M.~D.}\ \bibnamefont
  {Ventra}},\ }\bibfield  {title} {\bibinfo {title} {Transformer quantum state:
  a multipurpose model for quantum many-body problems},\ }\href@noop {}
  {\bibfield  {journal} {\bibinfo  {journal} {Physical Review B}\ ,\ \bibinfo
  {pages} {075147}} (\bibinfo {year} {2023})}\BibitemShut {NoStop}%
\bibitem [{\citenamefont {Wang}\ \emph {et~al.}(2022)\citenamefont {Wang},
  \citenamefont {Liu}, \citenamefont {Cheng}, \citenamefont {Liang},
  \citenamefont {Gu}, \citenamefont {Li}, \citenamefont {Ding}, \citenamefont
  {Jiang}, \citenamefont {Shi}, \citenamefont {Qian}, \citenamefont {Pan},
  \citenamefont {Chong},\ and\ \citenamefont {Han}}]{QuEst2022}%
  \BibitemOpen
  \bibfield  {author} {\bibinfo {author} {\bibfnamefont {H.}~\bibnamefont
  {Wang}}, \bibinfo {author} {\bibfnamefont {P.}~\bibnamefont {Liu}}, \bibinfo
  {author} {\bibfnamefont {J.}~\bibnamefont {Cheng}}, \bibinfo {author}
  {\bibfnamefont {Z.}~\bibnamefont {Liang}}, \bibinfo {author} {\bibfnamefont
  {J.}~\bibnamefont {Gu}}, \bibinfo {author} {\bibfnamefont {Z.}~\bibnamefont
  {Li}}, \bibinfo {author} {\bibfnamefont {Y.}~\bibnamefont {Ding}}, \bibinfo
  {author} {\bibfnamefont {W.}~\bibnamefont {Jiang}}, \bibinfo {author}
  {\bibfnamefont {Y.}~\bibnamefont {Shi}}, \bibinfo {author} {\bibfnamefont
  {X.}~\bibnamefont {Qian}}, \bibinfo {author} {\bibfnamefont {D.}~\bibnamefont
  {Pan}}, \bibinfo {author} {\bibfnamefont {F.}~\bibnamefont {Chong}},\ and\
  \bibinfo {author} {\bibfnamefont {S.}~\bibnamefont {Han}},\ }\bibfield
  {title} {\bibinfo {title} {Qu{E}st: Graph transformer for quantum circuit
  reliability estimation},\ }in\ \href@noop {} {\emph {\bibinfo {booktitle}
  {IEEE/ACM International Conference on Computer-Aided Design (ICCAD)}}}\
  (\bibinfo {year} {2022})\BibitemShut {NoStop}%
\bibitem [{\citenamefont {Nguyen}\ \emph {et~al.}(2024)\citenamefont {Nguyen},
  \citenamefont {Nguyen}, \citenamefont {Chen}, \citenamefont {Khan},
  \citenamefont {Churchill},\ and\ \citenamefont
  {Luu}}]{nguyen2024qclusformer}%
  \BibitemOpen
  \bibfield  {author} {\bibinfo {author} {\bibfnamefont {X.-B.}\ \bibnamefont
  {Nguyen}}, \bibinfo {author} {\bibfnamefont {H.-Q.}\ \bibnamefont {Nguyen}},
  \bibinfo {author} {\bibfnamefont {S.~Y.-C.}\ \bibnamefont {Chen}}, \bibinfo
  {author} {\bibfnamefont {S.~U.}\ \bibnamefont {Khan}}, \bibinfo {author}
  {\bibfnamefont {H.}~\bibnamefont {Churchill}},\ and\ \bibinfo {author}
  {\bibfnamefont {K.}~\bibnamefont {Luu}},\ }\bibfield  {title} {\bibinfo
  {title} {Q{C}lusformer: a quantum transformer-based framework for
  unsupervised visual clustering},\ }\href@noop {} {\bibfield  {journal}
  {\bibinfo  {journal} {arXiv preprint arXiv:2405.19722}\ } (\bibinfo {year}
  {2024})}\BibitemShut {NoStop}%
\end{thebibliography}
\end{document}